\documentstyle[12pt]{article}

\begin{document}


\hsize=6.15in
\vsize=8.2in
\hoffset=-0.42in
\voffset=-0.3435in

\normalbaselineskip=24pt\normalbaselines

\begin{center}
{\large \bf Model of the early development of thalamo-cortical
connections and area patterning via signaling molecules }
\end{center}

\vspace{0.15cm}

\begin{center}
{Jan Karbowski$^{1,*}$ and G.B. Ermentrout$^{2}$}
\end{center}

\vspace{0.05cm}

\begin{center}
$^{1}$ {\it  Sloan-Swartz Center for Theoretical Neurobiology,
Division of Biology 216-76, \\
California Institute of Technology,
Pasadena, CA 91125, USA } \\
$^{2}$ {\it Department of Mathematics,
University of Pittsburgh,
Pittsburgh, PA 15260, USA }
\end{center}


\vspace{0.1cm}

\begin{abstract}
The mammalian cortex is divided into architectonic and functionally
distinct areas. There is growing experimental evidence
that their emergence and development is controlled by both
epigenetic and genetic factors. The latter were recently implicated
as dominating the early cortical area specification. 
In this paper, we present a theoretical model that explicitly
considers the genetic factors and that is able to
explain several sets of experiments on cortical area regulation
involving transcription factors Emx2 and Pax6, 
and fibroblast growth factor FGF8. 
The model consists of the dynamics of thalamo-cortical connections
modulated by signaling molecules that are regulated genetically, and
by axonal competition for neocortical space.
The model can make predictions and provides a basic mathematical
framework for the early development of the thalamo-cortical connections
and area patterning that can be further refined as more experimental
facts become known. 
\end{abstract}





\noindent {\bf Running Title}: cortical patterning via signaling
molecules

\vspace{0.2cm}

\noindent $^{*}$ Corresponding author at:

\noindent Email: jkarb@cns.caltech.edu

\noindent Phone: (626)-395-5840

\noindent Fax: (626)-795-2397

\vspace{0.5cm}

\newpage

\section{Introduction}

Neocortex is organized into many functional subdivisions called areas
that have sharp boundaries. The areas can be identified anatomically
by investigating their distinct cytoarchitectonic properties and
unique connectivity patterns. One interesting feature is that the 
neocortical map composed of these areas is highly conserved within the
same species, and has common properties across different species
with different brain sizes (Nauta and Feirtag, 1986; Hofman, 1989;
Finley, 1995; Northcutt and Kaas, 1995; Krubitzer, 1995; 
Karbowski, 2003).

One of the main questions in the development of the mammalian cortex is
what factors control the specification and differentiation of
cortical areas. In the past, there were two opposing views.
One proposition was that areas are specified by intrinsic genetic
factors (Rakic, 1988) - the so-called protomap model. 
Another proposition was that areas are specified by extrinsic 
influence, i.e. by thalamo-cortical inputs (O'Leary, 1989). 
In recent years, however, the consensus has been growing that both
of these factors, i.e. genetic and epigenetic contribute to the
cortical area patterning (Krubitzer and Huffman, 2000; 
O'Leary and Nakagawa, 2002).
The genetic effects are thought to dominate the early stages of 
the development, while the epigenetic effects influence the later
stages. This view
has been stimulated by recent experiments demonstrating a direct
genetic involvement in cortical arealization (Bishop et al, 2000;
Mallamaci et al, 2000; Fukuchi-Shimogori and Grove, 2001, 2003; 
Bishop et al, 2002; Muzio et al, 2002; Garel et al, 2003).
In particular, it has been found that the genetic transcription
factors Emx2 and Pax6 control area specification even before
thalamo-cortical input arrives (Bishop et al 2000; Mallamaci et al
2000; Muzio et al, 2002; Bishop et al, 2002). Both of these factors 
are expressed
in a graded and complementary manner along antero-posterior (A/P)
axis in the neocortical ventricular zone. 
In mice with Emx2 mutation, anterior areas expand 
and posterior areas shrink, in Pax6 mutants the opposite is observed.
Another set of experiments has found that cortical maps can be
disrupted by modifying fibroblast growth factor FGF8 
(Fukuchi-Shimogori and Grove, 2001, 2003; Garel et al, 2003), 
which is also involved in embryonic patterning.
In wild type mouse neocortex, the FGF8 source is located in the
anterior pole giving rise to an expression concentration decaying
towards the posterior end. Under such regular conditions, the barrel field
(S1 area) is positioned in the center along A/P axis. However, increase
in the expression of FGF8 at the anterior pole, displaces the barrel
field more posteriorly. In contrast, blocking the FGF8 activity with
a soluble FGF8 receptor moves the field in the opposite direction.
Moreover, introduction of an extra source of FGF8 at the posterior
pole creates an additional barrel field that partly duplicates the
original field. Taken together, all these results suggest that
genetic perturbations can have a profound effect on the development
of thalamo-cortical connections and area patterning. Additionally,
these results are consistent with a hypothesis that the above signaling
molecules provide positional information for neuroepithelial cells
(Wolpert, 1969, 1996; Grove and Fukuchi-Shimogori, 2003).

Area-specific TC projections are probably controlled
by axon guidance molecules, similar to what happens in the retino-tectal
system (O'Leary et al, 1999). Although several molecules
have been found that are able to guide TC axons, there are some
experimental indications that the plausible candidates
for the late phase of accurate TC targeting are ephrins and their
receptors that are expressed both in the neocortex and in the thalamus
(Vanderhaeghen et al, 2000; Mackarehtschian et al, 1999; Mann et al,
2002; Uziel et al, 2002; Takemoto, 2002). It is well established that 
these molecules
direct axons to appropriate locations in many systems, in particular,
in the retino-tectal system (Goodhill and Richards, 1999), 
which, like the neocortex, preserves
the topography of projections.

The relationship between the axon guidance molecules and Emx2, Pax6,
and FGF8 is not known at present. It is likely, however, that the
latter control the expression level and gradients of the guiding
molecules. Thus, if the transcription factors and FGF8 concentrations
are modulated, then it should affect the concentration pattern of
the axon guidance molecules. This hypothesis is consistent with the
experiments on cortical areas shifting described above.

The purpose of this article is to provide a theoretical model of
the early TC projections and cortical area patterning.
The key assumption we make is that the signaling molecules Emx2,
Pax6, and FGF8 control the axon guidance molecules, which in turn
control TC pathfinding. We assume that there are three main types
of the axon guidance molecules, which we call A, B and C, that 
are expressed
in the neocortical ventricular zone in a graded and complementary
manner along A/P axis. In the model, these molecules repel and
attract different TC axons, which branch diffusively along the cortex, 
with specific forces, different for different classes of axons. 
Additionally, axons interact among themselves by competing for
neocortical space. As a consequence of these
interactions, a pattern of TC connectivity emerges that divides the
neocortex into regions with sharp boundaries defined by distinct TC
axon types. In this framework, area shifting experiments described
above can be understood as a result of shifts in the patterns of
expression of the signaling molecules.

\section{Model}

We assume that the early TC connectivity and cortical regionalization
along A/P axis is established by chemo-interaction between axons
originating from the thalamus and axon guidance molecules located on the
cortex (Sperry, 1963). In this paper, we assume that
there are $N=5$ main axon types. Under normal conditions,
 type $i=1$ corresponds to the axons coming from ventrolateral thalamus and
terminating in primary motor area M1, type $i=3$ corresponds to
the axons originating from thalamic ventrobasal complex and
terminating in S1, type $i=5$ relates to the axons connecting
thalamic LGN with the primary visual area V1. 
The remaining axons of type 2 and 4 terminate between areas M1 and
S1 (type $i=2$), and between S1 and V1 (type $i=4$), in order to ensure
a topographic TC connectivity.

In the model, it is assumed that there are three types of the axon
guidance molecules A, B, and C. In our choice of the number of guiding
molecules, we were motivated by experimental indications that
the best candidates
for the late phase of precise TC pathfinding are ephrins and their
receptors (O'Leary and Nakagawa, 2002; Lopez-Bendito and Molnar, 2003).
Up to now, 3 types of ephrins have been implicated as important:
ephrin-A5 expressed at high levels in the central part of the cortex,
ephrin-A4 expressed in the cortical intermediate zone with a gradient
decaying from the anterior towards the posterior end, and ephrin-B3
with a complementary gradient to ephrin-A4 (Lopez-Bendito and Molnar,
2003). In the model, the molecule A may mimic the action of ephrin-A4, 
the molecule B may act analogously to
ephrin-A5, and the molecule C may mimic ephrin-B3. The molecules A, B, and 
C have a mixed effect on TC axons, i.e. they both repel and attract them
and this is encoded in the interaction matrix $\gamma$. Experimentally,
ephrin-As and -Bs ligands have been shown to exhibit some selective
behavior, too. Ephrin-B expressing cells tend to attract axonal growth
cones with EphB receptors (McLaughlin et al, 2003), and repulse axons
with EphA4 receptors (Takemoto et al, 2002). The prevailing action 
of ephrin-As expressing 
cells is repulsion, which is well documented with axons expressing EphA
(e.g., Mann et al, 2002; McLaughlin et al, 2003).
However, when axonal cones express both EphA and ephrin-A as receptors,
then cells with ephrin-As can attract them (Knoll and Dreschner, 2002). 

In addition to the chemo-affinity, we assume that axons branch randomly
along A/P axis. Axonal branching has been found in the retino-tectal
(Yates et al, 2001) and retino-collicular (Simon and O'Leary, 1992) 
systems, and there are some indications (Mann et al, 2002) that similar 
effects are present in the 
establishment of TC connectivity. The basics of the process of random 
growth and decay of branches along axonal length can be mathematically
described by some stochastic component acting on a local branching
density (see below). The combined effect of the chemo-interactions 
and stochasticity on axonal terminals is mathematically equivalent to
their biased random growth on the cortical surface. The bias in growth
is caused by the gradients of the guidance molecules. For attractive
interactions, the larger the gradient the larger the growth rate. For
repulsive interactions, the larger the gradient the larger the decay rate.

Another crucial assumption is that there exists 
some sort of repulsive interaction 
between TC axons of different types that leads to their competition for
neocortical space. 
This constraint implies that the density of TC projections is limited,
because not all axon types are allowed to terminate 
simultaneously at a single point in the cortex. The idea of axon-axon 
competition was used 
in the past in the context of retino-tectal mapping (Prestige and
Willshaw, 1975; Fraser and Hunt, 1980) and it is consistent with
recent experiments in retino-collicular system (Feldheim et al, 2000).

In the mathematical model, the development of TC connections ($c_{i}$) 
between axons of type $i$ (with the branching density $a_{i}$) and cortical 
neurons $n$ is represented by a simple kinetic reaction scheme: 
$n + a_{i}\stackrel{\longrightarrow}{\leftarrow}^{\beta_{i}}_{\alpha_{i}}
c_{i}$ (e.g. Murray, 1993). Connections are destroyed
with the rate $\alpha_{i}$, and created with a rate proportional
to the product of the fraction of available neurons $n(x)$ and some 
power of branching density, i.e. $\beta_{i}n(x)[a_{i}(x)]^{k}$,
where $\beta_{i}$ is some positive constant. When the exponent $k > 1$, then
TC connections are created with a higher rate in regions with high 
concentration of axonal branches, so that there is some cooperativity. 
(In our numerical simulations we take $k=3$.)

 The full set of differential equations describing
the above processes is given by:

\begin{equation}
\frac{\partial{c_{i}(x)}}{\partial{t}} = -\alpha_{i} c_{i}(x)
+ \beta_{i} n(x)[a_{i}(x)]^{k},
\end{equation}
\begin{equation}
\frac{\partial{a_{i}(x)}}{\partial{t}} = \frac{\partial{J_{i}(x)}}
{\partial{x}}  + \alpha_{i} c_{i}(x) - \beta_{i} n(x)[a_{i}(x)]^{k},
\end{equation}
\begin{equation}
n(x,t) + \sum_{i=1}^{N} c_{i}(x,t) = 1,
\end{equation}\\
with the flux current

\begin{equation}
J_{i}(x)= D\frac{\partial{a_{i}}}{\partial{x}} 
- a_{i}(x)\left( \gamma_{Ai}\frac{\partial{\rho_{A}}}{\partial{x}}
+ \gamma_{Bi}\frac{\partial{\rho_{B}}}{\partial{x}} 
+ \gamma_{Ci}\frac{\partial{\rho_{C}}}{\partial{x}} \right),
\end{equation}\\
where $c_{i}(x)$ is a fraction of TC connections of type $i$ at position 
$x$ on the developing cortex, $n(x)$ is a fraction of cortical neurons at 
point $x$ available for a TC  connection (i.e. not already connected by
other TC axons), $a_{i}(x)$ is the branching density of axons of type $i$ 
at point $x$ on the cortex that originate from the thalamus, and $N$ 
is the number of axon types. Our model is a population model, which deals 
with densities and fractions, and does not take directly into account 
the fine structure of axonal branching on a single axon level, such as
complex arborization patterns; instead we consider population branching 
density. (For detailed modeling of branching structures and their influence 
on the retino-topic map, see Yates et al, 2004).
Eq. (3) is a mathematical  consequence of axonal competition
for neocortical space, and it provides
a conservation law for the fractions of available (unconnected) neurons 
and all the connection types for every point on the cortex and for all times.
The flux $J_{i}(x)$ is associated with the growth and decay of axonal 
branches of type $i$ at point $x$. It is composed of the following 
contributions: the diffusive axonal branching that is proportional 
to the diffusion constant $D$, and chemo-interaction between 
axonal branches and the axon guidance molecules that have concentrations 
$\rho_{A}(x)$, $\rho_{B}(x)$, and $\rho_{C}(x)$ (Sperry, 1963). 
The nature of this interaction is additive, that is, the guiding molecules
do not interact with themselves, and it can be either repulsive or attractive,
which is controlled by the sign and strength of the parameters 
$\gamma_{Ai}$, $\gamma_{Bi}$, and $\gamma_{Ci}$.

Our model focuses on how the anterior/posterior boundaries of cortical
areas are formed. Thus, from a theoretical point of view, this
biological problem has a one-dimensional character and
the spatial variable $x$ measures distances along
the A/P axis. In our model, the anterior pole corresponds to $x\approx 0$,
and the posterior pole to $x\approx L$, where $L$ is the linear
size of the cortex in that direction.

We adopt  sealed-end boundary conditions for the flux current $J_{i}(x)$, 
i.e. $J_{i}(L)= J_{i}(0)= 0$, since in our system axonal branches cannot 
grow outside the boundaries of the cortex. This choice implies
that the total number of axonal  branches 
and connections of any given type in the system is constant, i.e. 
$\int_{0}^{L} dx [a_{i}(x)+c_{i}(x)]$ is time independent (this follows
from adding eqs. (1) and (2)). Thus initially, there are many branches
but no connections. However, as time progresses, the total number of
branches decreases at the expense of the formed connections. In a real
biological system, however, this constraint may be satisfied only 
approximately.

It is assumed that the axon guidance molecules located on the cortex are
regulated by the transcription factors Emx2 and Pax6, and the fibroblast
growth factor FGF8. Recent experiments (Fukuchi and Grove 2003; Garel 
et al, 2003) show that
(i) FGF8 and Emx2 mutually inhibit each other. The Fukuchi and Grove
(2003) results also suggest that (ii) FGF8 directly
controls the location of cortical maps, and additionally that
(iii) Emx2 indirectly controls the positioning of cortical maps by 
acting upstream of FGF8 and regulating it. 
Another set of experiments (Muzio et al, 2002) shows that (iv) the two
transcription factors, Emx2 and Pax6, mutually repress each other.
Based on these facts, one can construct a minimal signaling pathway
(Fig. 1) that is responsible for regulating expression levels of the
guiding molecules A, B, and C. In this model, both Emx2 and Pax6 act
upstream of FGF8, but only Emx2 regulates it directly. 
FGF8 and Pax6 have similar gradients, i.e. their concentrations
decay from the anterior towards the posterior end, while Emx2
has an opposite gradient to them. 
The output signal $f$ from the pathway can be viewed as a renormalized
(by interactions with other signaling molecules) concentration of FGF8. 
We assume that either this signal or some other $f$-activated signal
serves as a morphogen
signal (Grove and Fukuchi-Shimogori, 2003) that provides a positional 
information for cortical cells, similar to what  happens in other 
developing systems (Wolpert, 1969, 1996).
In response to the signal $f$, cells express different levels of the
guiding molecules, which in turn control TC axons.

The diagram in Fig. 1 can be described mathematically in the
following way.
The transcription factors Emx2 and Pax6 both repress each other
with  strengths $v_{1}$ and $v_{2}$, respectively ($v_{1}, v_{2} > 0$). 
Emx2 negatively controls FGF8 with a strength $w_{1}$, and FGF8 
in turn inhibits Emx2 with a strength $w_{2}$ ($w_{1}, w_{2} > 0$).  
From all of these relationships, it follows that FGF8 ($f$)
is regulated negatively by Emx2 and positively by Pax6. 
Mathematical equations describing the dynamics of the expression of Emx2, 
Pax6, and FGF8 are given by

\begin{equation}
\tau_{s}\frac{ds(x)}{dt}= - s(x) +
\frac{\eta_{emx}(x)}{1+w_{2}f(x)+v_{2}r(x)}
\end{equation}
\begin{equation}
\tau_{r}\frac{dr(x)}{dt}=  - r(x) +
\frac{\eta_{pax}(x)}{1+v_{1}s(x)}
\end{equation}
\begin{equation}
\tau_{f}\frac{df(x)}{dt}=  - f(x) +
\frac{\eta_{fgf}(x)}{1+w_{1}s(x)},
\end{equation}\\
where $s(x)$, $r(x)$, and $f(x)$ are expression levels (renormalized
concentrations) of Emx2, Pax6, and FGF8, respectively, in the presence
of interactions between these molecules.
The parameters $\tau_{s}$, $\tau_{r}$, and $\tau_{f}$ 
are signaling time constants, and finally $\eta_{emx}(x)$, 
$\eta_{fgf}(x)$, $\eta_{pax}(x)$ are uncoupled (i.e. without interactions)
concentrations of Emx2, FGF8, and Pax6. 
We assume the following
forms of the uncoupled concentrations for Emx2, Pax6, and FGF8:
$\eta_{emx}(x)= A_{emx} \exp[-(x-L)^{2}/\zeta_{emx}^{2}]$,
$\eta_{pax}(x)= A_{pax} \exp[-x^{2}/\zeta_{pax}^{2}]$, and
$\eta_{fgf}(x)= A_{fgf} \exp[-x^{2}/\zeta_{fgf}^{2}]$, where
$A_{emx}$, $A_{pax}$, $A_{fgf}$ are amplitudes of expression, and
$\zeta_{emx}$, $\zeta_{pax}$, $\zeta_{fgf}$
are constants characterizing the ranges of expression with values in 
the interval $(0,L)$.
When two sources of FGF8 are present, one in the anterior and
second in the posterior end, then the uncoupled concentration
of FGF8 takes the form:
$\eta_{fgf}(x)= A_{fgf} \exp[-x^{2}/\zeta_{fgf}^{2}] +
 A_{fgf}' \exp[-(x-L)^{2}/\zeta_{fgf}'^{2}]$, where
$A_{fgf}'$ and $\zeta_{fgf}'$ are the amplitude and the range of
the posterior source, respectively.
These uncoupled
concentrations act as signals produced at boxes denoted
by Emx, FGF, and Pax in Fig. 1. We assume that the time constants
$\tau_{s}$, $\tau_{r}$, and $\tau_{f}$ are much smaller than 
characteristic time
constants associated with the processes involving TC projections.
That is, at the time when TC axons arrive to the cortical surface,
all these molecules have already reached their steady-state, which is
consistent with experimental data (Cohen-Tannoudji et al, 1994;
Miyashita-Lin et al, 1999; Nakagawa et al, 1999).
For this reason, we are interested in a steady-state
solution for $f(x)$. 
Its plot is displayed in Fig. 2a.
Under normal conditions, its spatial profile displays a monotonic decay
from the anterior towards the posterior end (Fig. 2a), i.e. it has
a similar gradient to the uncoupled FGF8.

We assume
that the stationary distributions of the guiding molecules are
as follows: the molecule A is mostly expressed in those regions
of the cortex for which the signal $f(x)$ is high (anterior end),
B is expressed mostly in regions where $f(x)$ is moderate (center),
and C has high concentration in locations where $f(x)$ is weak
(posterior end). These relationships can be represented mathematically
in the following way:

\begin{equation}
\rho_{A}(x)= G_{A}\left(f(x)-\theta_{1}\right),
\end{equation}
\begin{equation}
\rho_{B}(x)= G_{B}\left(\theta_{2}-f(x)\right)
G_{B}\left(f(x)-\theta_{3}\right),
\end{equation}
\begin{equation}
\rho_{C}(x)= G_{C}\left(\theta_{4}-f(x)\right),
\end{equation}\\
where $\rho_{A}$, $\rho_{B}$, and 
$\rho_{A}$ are stationary concentrations of the guiding
molecules A, B, and C. The thresholds
$\theta_{i}$ $(i=1,...,4)$ are some positive constants such
that $\theta_{4} < \theta_{3} < \theta_{2} < \theta_{1} < f(0)$, and
$G_{j}(y)= (\kappa_{j}/2)[1 + \tanh(y/\sigma_{j})]$, where
$\kappa_{j}$ and $\sigma_{j}$  (for $j=A, B, C$) are some positive
constants controlling the amplitude and the slope of the
concentration $\rho_{j}(x)$. The parameters $\sigma_{j}$ are
chosen such as to ensure that the expressed levels of the guiding
molecules A, B, and C have graded concentrations (Fig. 2b).

\section{Results}

All the results reported in this section were obtained by numerically solving
 eqs. (1)-(10). We used a second-order Runge-Kutta method
for the ordinary differential equations and a Crank-Nicholson method
(second order accurate in time and space) for the partial 
differential equations. The values of the parameters used are given
in the legend of Fig. 2.

\noindent {\bf Normal area positioning.}

Under normal conditions (wild type) cortical areas in the mouse neocortex 
are located such that motor area M1 occupies anterior part, sensory area
S1 occupies the central part, and visual area V1 is positioned in the
posterior end. Areas are defined as a spatial
pattern of TC connectivity fraction $c_{i}(x)$ for $i=1,...,5$. 
A high value of $c_{1}(x)$ corresponds to the M1 field, a high value of 
$c_{3}(x)$ corresponds to the S1 area, and a high value of $c_{5}(x)$
corresponds to V1. For the remaining two connection types $c_{2}(x)$
and $c_{4}(x)$, their high values correspond to areas between M1
and S1, and between S1 and V1, respectively. Both, the connectivity 
fraction $c_{i}(x,t)$ and the axonal branching density $a_{i}(x,t)$ evolve 
in time to a steady-state according to eqs. (1)-(4) (see Figs. 3 and 4). 
Initial conditions are chosen such that
$a_{i}(x,t=0)$ is uniformly (although with some noise) distributed
in space, and $c_{i}(x,t=0)= 0$, for $i=1,...,5$. This choice is motivated
by experimental data (Simon and O'Leary, 1992; Yates et al, 2001), 
suggesting that the early TC connectivity in mammals is established in
three phases. The first phase is axonal overshooting along A/P axis 
beneath the cortical plate. The second phase is composed of axonal 
branching along their length, and the final phase provides stabilization 
of topographically correct axonal collaterals and elimination of distant 
branches. Our modeling starts from the second phase. As time progresses
both of the above distributions evolve into spatially heterogeneous
state with a close relationship between patterns of $a_{i}(x)$
and $c_{i}(x)$ (compare Figs. 3c and 4c). The reason for this relationship
can be disclosed if we use eqs. (1) and (3). Then one can
derive a formula relating the stationary distribution of the connections 
$\overline{c_{i}}(x)$ with the stationary branching densities 
$\overline{a_{i}}(x)$:

\begin{equation}
\overline{c_{i}}(x)= \frac{\beta_{i}[\overline{a_{i}}(x)]^{k}}{\alpha_{i} 
+ \sum_{j=1}^{N} \beta_{j}[\overline{a_{j}}(x)]^{k} }.
\end{equation}\\
From this equation, it follows that the area of type $i$ emerges in regions
with high concentration of the branches of type $i$. However, one should
note that not all diffusive branches form connections.
This can be seen in Fig. 3c, where branches of type 2 and 4 have rather
broad distributions, which are absent in the connectivity pattern in
Fig. 4c.

The locations of cortical regions are shaped
by different forces coming from the axon guidance molecules A, B and C.
These forces activate the growth of axonal branches in some locations
and inhibit their growth in other locations.
To have, at a steady-state, a high density of axons of type $i=1$ in the 
anterior end, it is assumed that they are moderately/strongly attracted
by the molecules A, 
repulsed strongly by the molecules C, and repulsed moderately by B.  
Similarly, to have high density of
axons of type $i=5$ in the posterior end, it is assumed that they are
moderately/strongly attracted by C, strongly repulsed by the molecules A, 
and moderately repulsed by B.
Axons of type $i=3$ are forced to aggregate in the center by assuming
that they are attracted by B, and repulsed, either strongly or moderately, 
with approximately the same strength from the guiding molecules A and C. 
Elements of the interaction matrix $\gamma_{\alpha i}$ for axons of type 
$i=2$ and $i=4$ are all negative (repulsion),
and are chosen such that these axons
terminate in between axons 1, 3 and 3, 5 in the wild type case.
Although the concentrations of A, B, and C are graded (Fig. 2b), 
the spatial pattern of the fractions of TC connections $c_{i}(x)$ is almost
exclusive with sharp borders (Fig. 4c), which leads to the emergence of
cortical areas (Fig. 4d). The exclusiveness comes from the
axonal competition constraint we imposed (see eq. (3)).
The width of the areal border
is directly proportional to the diffusion constant $D$, and inversely
proportional to the product of the coefficient $k$ and
the magnitude of the interaction between guiding molecules and axons. 
Purely repulsive
interactions increase the border width in comparison to cases with mixed 
repulsive/attractive interactions (see Fig. 9 in the discussion).

\noindent {\bf Area shifting.}

In the area shifting experiments expression levels of Emx2, Pax6, and
FGF8 were affected 
(Bishop et al, 2000; Mallamaci et al, 2000;
Fukuchi-Shimogori and Grove, 2001; Bishop et al, 2002; Muzio et al
2002; Garel et al 2003, Fukuchi-Shimogori and Grove, 2003).
We model this effect by reducing or amplifying
the appropriate concentration amplitude ($A_{emx}, A_{pax}, A_{fgf}$)
and the range of expression
of one of these molecules. In Fig. 5, we present posteriorly shifting 
areas as we decrease the amplitude $A_{emx}$ to zero. This case corresponds
to loss-of-function Emx2 mutants for which such shifting has been
reported (Bishop et al, 2000; Mallamaci et al, 2000; Muzio et al, 2002;
Bishop et al, 2002). 
The interpretation of this effect in our model
is as follows. Reducing the concentration of Emx2 effectively increases 
the output signal $f$ (see, Fig. 1), since then the Emx2 $\rightarrow$
FGF8 inhibition decreases.
This leads  effectively to shifting posteriorly the expression profile
of $f$ (Fig. 5a), and consequently shifting posteriorly the expression pattern
of the guiding molecules (Fig. 5b).
That new pattern biases axonal branching growth to the right to new 
locations, where balanced forces act on them (Fig. 5c).
Again, there is a close relationship between distributions of axons and
TC connectivity patterns (Figs. 5c,d). All areas are shifted posteriorly
(Fig. 5e). In particular, a clear shift in the position of 
the S1 area can be seen by comparing figures 4c and 5d.

Fig. 6 displays anteriorly shifting areas as we decrease the amplitude
$A_{pax}$ to zero. This case corresponds to loss-of-function Pax6 mutants
for which shifting in this direction has been experimentally 
reported (Bishop et al, 2000; Muzio et al, 2002; Bishop et al, 2002).  
By the same token, reduction in Pax6 concentration decreases
effectively the output signal $f$ (Fig. 6a), which in turn shifts
anteriorly the concentrations of the molecules A, B and C (Fig. 6b). 
This rearrangement pushes TC axons more into anterior end to new 
equilibrium positions (Fig. 6c) that leads to areas shift in this
direction (Figs. 6d,e).

The action of FGF8 is only slightly different. When we increase
the expression level of FGF8 by amplifying its amplitude $A_{fgf}$
and its range, we also increase the signal $f$ that leads to posterior 
shift in expression levels of the molecules A, B and C (Fig. 7a). 
As a consequence of this, TC connectivity (Fig. 7b) and cortical
areas (Fig. 7c) shift posteriorly, similarly to what happens in the case
of reduced Emx2. In contrast, if we decrease the expression of
FGF8 by decreasing $A_{fgf}$ and the range of its expression, 
then we obtain pattern shifting in the opposite direction by the same 
mechanism as above (Figs. 7d,e,f). Both types of shifting have been 
experimentally observed  recently (Fukuchi-Shimogori and Grove, 2001, 2003; 
Garel et al, 2003).

\noindent {\bf Generation of two separate S1 fields.}

In experiments with an extra source of FGF8 at the posterior end,
in some cases, a second entirely separate S1 barrel field can be
generated (Fukuchi-Shimogori and Grove, 2001).
We can mimic this effect of ``mirror symmetry'' in our model (Figs. 8a,b). 
The key observation is that by modifying the expression level of FGF8,
one changes the spatial shape of the signal $f(x)$. Without the
second ectopic source of FGF8 at the posterior end, this function
decays monotonically from the anterior to the posterior end. However,
when this second source is present, the function $f(x)$ can
have a minimum at central part under some conditions (Fig. 8c). 
This happens if the diffusion range of the ectopic FGF8 is not too
large and if its amplitude is sufficiently strong to overcome
inhibition of Emx2 in the posterior end. From eq. (9), then it follows
that $\rho_{B}(x)$ is expressed at high levels at two separate
central locations which partly overlap (Fig. 8d). Thus, instead of
growing to one central location, axons of type $i=3$ grow
to two slightly separated locations that correspond to
two partly symmetric S1 areas (Figs. 8a,b). However, we are unable, within
our approach, to distinguish between two possibilities: one, that the
same group of axons target two S1 fields by branching in two separate
locations, and second, that axons subdivide into two groups, each
targeting only one S1 field. These two possibilities await experimental
teasing apart.

Surprisingly, this is not the only modification that the second
ectopic source of FGF8 brings. It turns out that the whole pattern
of both TC connectivity (Fig. 8a) and cortical fields (Fig. 8b) 
has a mirror symmetry.
Area V1 ($c_{5}$) is positioned in the gap between the two S1 
($c_{3}$) areas, and area M1 ($c_{1}$) has two branches:
one in the regular anterior end, and another in the posterior
end that under normal conditions would be occupied by V1. Also,
area $c_{2}$ has two branches, and area $c_{4}$ overlaps S1 and V1.
These results can be understood by invoking again Figs. 8c,d and 
eqs. (8)-(10). The guiding molecule C that acts as an attractant for
axons terminating in V1 area is expressed now only slightly in the central
part due to the modified shape of the signal $f(x)$ (only for
$x$ located in the center, $f(x)$ is relatively close to the threshold
$\theta_{4}$). Axonal branches of type $i=5$ gathered in the center due 
to attraction to C, are additionally repulsed by two regions of high 
concentration of B (Fig. 8d), and that effectively pushes them to the gap 
between the two S1 fields. Similar arguments hold for the M1 ($c_{1}$) 
fields. The molecule A that attracts axons of type 1 terminating
in M1 is expressed now not only in the anterior end but also partially
in the posterior end (Fig. 8d). This leads to the type $i=1$ axons
terminating in each of these locations (Fig. 8a). 
Similarly for the area $c_{2}$.
Axonal branches of this type are the most repulsed by the molecules B and C, 
and only weakly repulsed by A. As a result, they grow in the regions
where the molecule A has a moderate expression, that is, also in two
symmetric locations.

From the above it is apparent that the ectopic FGF8 can profoundly
affect the architecture of the neocortex by creating and destroying
areas at different locations. It would be interesting to verify
these predictions experimentally.

\section{Discussion}

In this paper, a mathematical model of the early development
of TC connections and cortical area patterning is presented. 
Cortical patterning is achieved in the model by allowing TC axonal
branches to undergo a combination of biased random growth with their mutual 
competition for neocortical space.
The model captures the essential components, which have
been experimentally implicated as the important genetic factors 
(Bishop et al, 2000; Mallamaci et al, 2000;
Fukuchi-Shimogori and Grove, 2001, 2003; Bishop et al, 2002; Muzio et al
2002; Garel et al 2003).
Its strength is in its ability to reproduce several recent 
experiments within a single theoretical framework, 
and also in its potential to predict the outcomes of some experiments.
In this regard, one of the tests of this model can be experimental
verification of the theoretical findings related to duplication of S1
field. In particular, to test (i) if cells in the gap between two
symmetric S1 fields really acquire the properties typical for neurons in
the visual cortex, and (ii) if cells in the
posterior end acquire properties typical for motor neurons.

Other tests of the model that are perhaps more feasible experimentally
could include the influence of the transcription factors Emx2 and Pax6, 
and FGF8 on the guidance molecules. If the guiding molecules A, B, and C 
really correspond to ephrins A4, A5, and B3, respectively, then by 
manipulating Emx2, Pax6, and FGF8 one should be able to observe changes
in ephrins concentrations. Specifically, in shifting experiments,
we predict that for Emx2 mutants the concentrations of all three ephrins
should shift posteriorly (Fig. 5b), and for Pax6 mutants they should
shift anteriorly (Fig. 6b) as compared to the wild-type case. Partial 
confirmation of this prediction is provided by the data of Bishop et al 
(2002), who found that ephrin-A5 concentration shifts posteriorly in Emx2 
mutants and it shifts anteriorly in Pax6 mutants. In the case of FGF8, 
we predict that when FGF8 is overexpressed in the anterior pole, 
then all ephrins should shift posteriorly (Fig. 7a), and when FGF8 is 
underexpressed, they should shift in the opposite direction (Fig. 7d).
In the case of two sources of FGF8 present at the two poles, a crucial
experiment would be to verify Fig. 8d. In particular, the model predicts
that the molecule A (ephrin-A4) should be expressed not only in the
anterior end (regular location) but also in the posterior end, and 
additionally that the molecule C (ephrin-B3) should be minimally expressed 
in the center of the cortex instead of the posterior end. Such experiment 
would prove or disprove the correctness of our assumption on the relationship
between FGF8 and the axon guidance molecules.

Our model is robust to changes in the interaction parameters between
transcription factors ($v_{1}, v_{2}, w_{1}, w_{2}$). However, these
parameters cannot be too small, since then the effects of area shifting
are small too. In general, the larger these parameters the more pronounced
the shiftings.

The most crucial interactions in the model are the chemo-interactions 
between the axon guidance molecules and TC axons, i.e. the matrix $\gamma$.
Together with axonal competition (eq. (3)), the elements of this matrix 
control the location and border sharpness of 
cortical areas. The results do not depend dramatically on their precise
values and selectivity in sign, but rather they depend on their relative
magnitudes. In order to obtain the correct TC projections in wild-type
conditions, it is essential to have the following pattern in the 
chemo-interactions: axons that are supposed to terminate in the anterior
part of the cortex should be strongly repulsed by the guidance molecules
located in the posterior end, moderately repulsed by the centrally located
guidance molecules, and either attracted or very weakly repulsed by the
anterior guidance molecules. Analogically, for the axons that are supposed
to terminate in the posterior part, the reverse pattern of interactions
is necessary, i.e. these axons should be strongly repulsed by the anterior
guidance molecules, moderately repulsed by the central guidance molecules,
and either attracted or very weakly repulsed by the posterior molecules.
For the axons that are supposed to terminate in the central part of the
cortex, it is important that they interact with the posterior and anterior
guidance molecules in a symmetric manner, i.e. they should be repulsed
by them with approximately the same strength. The larger this strength
the more confined the axonal projections are to the central part. The
interaction with the central guidance molecules is not very important,
i.e. it can be attractive (of any strength) or repulsive (but the strength
cannot exceed the repulsion from the lateral molecules).

In this paper we assumed that the axon guidance molecules act selectively
such that they repulse some classes of TC axons and attract other. However,
this assumption is not crucial for the results. Even, if it turns out that
the net effect of chemo-interaction between cortical ephrins 
(and possibly other guidance molecules) and TC axons is repulsion,
the appropriately modified
interaction matrix would not qualitatively change the results presented 
here. As an example, in Fig. 9 we show a spatial
profile of TC connectivity when the matrix $\gamma$ has only negative
elements but with the pattern of their inter-relationships discussed above.
In this case one can still generate separate areas, although
with much less sharp borders. They shift as Emx2, Pax6 or FGF8 are up- or
downregulated, similarly as it happens in the case with the interaction 
matrix that has negative and positive elements (Fig. 9b). Also, one
can generate a mirror symmetry effect when two sources of FGF8 are
present at two separate poles (Fig. 9c).

It is interesting to note that an analog  of the mirror symmetry effect
occurs also in the retino-tectal system (Fig. 2c,d in McLaughlin et al,
2003). This takes place in mice with ephrin-A5 and ephrin-A2/ephrin-A5
mutants. There is some analogy with our case of the double FGF8 source, 
because in our model its second ectopic source also modifies
the guiding molecules, especially their posterior distributions.

In the model we consider, motivated by experimental data, the three types
of the guidance molecules. However, one can obtain qualitatively similar
results with only one type of guiding molecule located primarily either
in the posterior or anterior part of the cortex. Regionalization of the
cortex then would be obtained by appropriately choosing the elements of
the matrix $\gamma$ in the similar way as we did in this paper. The seemingly
uneconomical number of the axon guidance molecules chosen by nature may be
necessary for a finer targeting of TC axons within cortical areas.

Our model is the first study, which takes into account genetic factors
(Emx2, Pax6, FGF8) and their influence on the gradients of the guiding
molecules. However, some aspects of the axon guidance part have been
modeled previously in the context of retino-tectal projections 
(Prestige and Willshaw, 1975; Fraser and Hunt, 1980;
Whitelaw and Cowan, 1981; Gierer, 1983). The model of Whitelaw and Cowan
(1981) puts a greater emphasis on synaptic/neural activity than on
guidance molecules. The model of Gierer (1983) seems to be somewhat abstract; 
it uses minimization procedure of some abstract function related to
retinal and tectal gradients in order to obtain topographic mapping. 
The axonal guidance part of our model is 
closest in spirit to the models of Prestige and Willshaw (1975), and
Fraser and Hunt (1980), which also used the idea of axonal competition. 
However, there are several significant differences between these models
and our model. Our main objective was to generate topographic arealization
with sharp borders, while their goal was to generate continuous topography
without areas. In terms of mathematics, our model uses the reaction 
kinetics equations (1-4) on a population level, while Prestige and Willshaw
(1975) use algorithmic approach with a discrete time on a single axon/neuron 
level. Fraser and Hunt (1980), on the other hand, impose several constraints 
on axonal dynamics that guide them to the correct locations. Also,
we consider explicitly the forces between TC axons and the guidance molecules,
and additionally stochastic branching effects employed by the diffusion term 
(see, Eq. (4)); these features are absent in their models.

The present model can be also modified by including more biophysical details,
more signaling molecules in the pathway, more pathways, and 
cross-interactions between pathways, if necessary. 
For example, it is likely that more signaling molecules
are involved in regulating axon guidance molecules and that the simple
scheme in Fig. 1 will be expanded as more experimental data becomes
available. Other candidates playing a role similar to FGF8 but providing
positional information from another cortical end, could be
WNT and BMP molecules located on the cortical hem (O'Leary and Nakagawa,
2002; Grove and Fukuchi-Shimogori, 2003).
However, if those molecules act in coordination with FGF8, then
it is unlikely that such an expanded model would
qualitatively change the present results.

Throughout the paper, we assume that genetic effects determine the very
early stage of cortical development and set the basic parcelation of
the neocortex (Miyashita-Lin et al, 1999; Nakagawa et al, 1999).
Thus, we model only the early cortical development,
before and right after the arrival of TC input. Later stages of the
development are probably controlled by TC input (i.e. molecules diffused
by arrived TC axons), and by activities of neurons and their synapses
(Katz and Shatz, 1996), and therefore require a different approach. 
However, it is likely that those late stage activities only refine that basic
patterning plan set by genetic factors 
(O'Leary and Nakagawa, 2002; Lopez-Bendito and Molnar, 2003).
In this respect, it is probable that neural activities can additionally
reduce a partial overlap of the areas that is due to molecules diffusion
(e.g. see Fig. 4c). Also those axons/branches diffusing around the cortex that
do not establish connections, will likely die out at later stages of the
development. That effect is not included in our model.

\noindent{\bf Acknowledgments}

We thank Elizabeth Grove for comments on a draft of this paper.
The work was supported by the Sloan-Swartz fellowship at Caltech (J.K.) 
and the National Science Foundation (G.B.E.).

\vspace{1.5cm}

\noindent {\bf References} 

\noindent Bishop, K.M., Goudreau, G., and O'Leary, D.D. (2000).
Regulation of area identity in the mammalian neocortex by Emx2 and
Pax6. Science  288, 344-349.

\noindent Bishop, K.M., Rubenstein, J.L.R., and O'Leary, D.D.
(2002). Distinct actions of Emx1, Emx2 and Pax6 in regulating the
specification of areas in the developing neocortex. J. Neurosci.
22, 7627-7638.

\noindent Cohen-Tannoudji, M., Babinet, C., and Wassef, M. (1994).
Early determination of a mouse somatosensory cortex marker.
Nature 368, 460-463.

\noindent Feldheim, D.A., et al. (2000). Genetic analysis of ephrin-A2
and ephrin-A5 shows their requirement in multiple aspects of 
retinocollicular mapping. Neuron, 25, 563-574.

\noindent Finlay, B.L., and Darlington, R.B. (1995). Linked regularities
in the development and evolution of mammalian brains. Science
268, 1578-1584.

\noindent Fraser, S.E., and Hunt, R.K. (1980). Retinotectal specificity:
models and experiments in search of a mapping function. Annu. Rev.
Neurosci. 3, 319-352.

\noindent  Fukuchi-Shimogori, T., and Grove, E.A. (2001). Neocortex
patterning by the secreted signaling molecule FGF8. Science 
294, 1071-1074.

\noindent  Fukuchi-Shimogori, T., and Grove, E.A. (2003). 
Emx2 patterns the neocortex by regulating FGF positional signaling.
Nature Neurosci. 6, 825-831.

\noindent  Garel, S., Huffman, K.J., and Rubenstein, J.L.R. (2003).
Molecular regionalization of the neocortex is disrupted in FGF8
hypomorphic mutants. Development  130, 1903-1914.

\noindent  Gierer, A. (1983). Model for the retino-tectal projection.
Proc. R. Soc. Lond. B 218, 77-93.

\noindent  Goodhill, G.J., and Richards, L.J. (1999). Retinotectal
maps: molecules, models and misplaced data. Trends Neurosci. 22, 529-534.

\noindent  Grove, E.A., and Fukuchi-Shimogori, T. (2003). Generating
the cerebral cortical area map. Annu. Rev. Neurosci. 26,
355-380.

\noindent Hofman, M.A. (1989). On the evolution and geometry of 
the brain in mammals. Prog. Neurobiol. 32, 137-158.

\noindent Karbowski, J. (2003). How does connectivity between 
cortical areas depend on brain size? Implications for efficient computation.
J. Comput. Neurosci. 15, 347-356.

\noindent  Katz, L.C., and Shatz, C.J. (1996). Synaptic activity
and the construction of cortical circuits. Science 274,
1133-1138.

\noindent Knoll, B., and Drescher, U. (2002). Ephrin-As as receptors
in topographic projections. Trends Neurosci. 25, 145-149.

\noindent Krubitzer, L. (1995). The organization of neocortex in mammals:
Are species differences really so different? Trends Neurosci. 
18, 408-417.

\noindent  Krubitzer, L., and Huffman, K.J. (2000). Arealization 
of the neocortex in mammals: genetic and epigenetic contributions to the 
phenotype. Brain Behav. Evol. 55, 322-335.

\noindent  Lopez-Bendito, G., and Molnar, Z. (2003). Thalamo-cortical
development: How are we going to get there? Nature Rev. Neurosci.
4, 276-289.

\noindent  Mackarehtschian, K., Lau, C.K., Caras, I., and
McConnel, S.K. (1999). Regional differences in the developing
cerebral cortex revealed by $Ephrin-A5$ expression. Cereb.
Cortex 9, 601-610.

\noindent Mallamaci, A., Muzio, L., Chan, C.H., Parnavelas, J.,
and Boncinelli, E. (2000). Area identity shifts in the early cerebral
cortex of $Emx2^{-/-}$ mutant mice. Nature Neurosci. 3,
679-686.

\noindent  Mann, F., Peuckert, C., Dehner, F., Zhou, R.,
and Bolz, J. (2002). Ephrins regulate the formation of terminal
axonal arbors during the development of thalamocortical projections.
Development 129, 3945-3955.

\noindent McLaughlin, T., Hindges, R., and O'Leary, D.D.M. (2003).
Regulation of axonal patterning of the retina and its topographic
mapping in the brain. Curr. Opin. Neurobiol. 13, 57-69.

\noindent Miyashita-Lin, E.M., Hevner, R., Wassarman, K.M.,
Martinez, S., and Rubenstein, J.L.R. (1999). Early neocortical
regionalization in the absence of thalamic innervation. Science 
285, 906-909.

\noindent  Murray, J.D. (1993). {\it Mathematical Biology. \/}
Springer-Verlag, Berlin, pp. 109-139.

\noindent Muzio, L. et al (2002). Emx2 and Pax6 control 
regionalization of the pre-neuronogenic cortical primordium. 
Cereb. Cortex 12, 129-139.

\noindent Nakagawa, Y., Johnson, J.E., and O'Leary, D.D.M. (1999).
Graded and areal expression patterns of regulatory genes and cadherins
in embryonic neocortex independent of thalamocortical input.
J. Neurosci. 19, 10877-10885.

\noindent Nauta, W.J.H., and Feirtag, M. (1986). {\it 
Fundamental Neuroanatomy. \/} (Freeman, New York).

\noindent Northcutt, R.G., and Kaas, J.H. (1995). The emergence
and evolution of mammalian neocortex. Trends Neurosci. 
18, 373-379.

\noindent  O'Leary, D.D.M. (1989). Do cortical areas emerge from
a protocortex? Trends Neurosci. 12, 400-406.

\noindent  O'Leary, D.D.M., Yates, P.A., McLaughlin, T. (1999).
Molecular development of sensory maps: representing sights and smells
in the brain. Cell 96, 255-269.

\noindent  O'Leary, D.D.M., and Nakagawa, Y. (2002). Patterning
centers, regulatory genes and extrinsic mechanisms controlling
arealization of the neocortex. Curr. Opin. Neurobiol. 12,
14-25.

\noindent Prestige, M.C., and Willshaw, D.J. (1975). On a role for
competition in the formation of patterned neural connexions.
Proc. Roy. Soc. Lond. B 190, 77-98.

\noindent Rakic, P. (1988). Specification of cerebral cortical
areas. Science 241, 170-176.

\noindent Simon, D.K., and O'Leary, D.D.M. (1992). Development of
topographic order in the mammalian retinocollicular projection.
J. Neurosci. 12, 1212-1232.

\noindent  Sperry, R.W. (1963). Chemoaffinity in the orderly
growth of nerve fibers patterns and connections. Proc.
Natl. Acad. Sci. USA  50, 703-710.

\noindent  Takemoto, M., et al (2002). Ephrin-B3-EphA4
interactions regulate the growth of specific thalamocortical
axon populations in vitro.  Eur. J. Neurosci. 16, 1168-1172.

\noindent  Uziel, D. et al (2002). Miswiring of limbic 
thalamocortical projections in the absence of ephrin-A5. 
J. Neurosci. 22, 9352-9357.

\noindent  Vanderhaeghen, P. et al (2000). A mapping label
required for normal scale of body representation in the cortex.
Nature Neurosci. 3, 358-365.

\noindent  Whitelaw, V.A., and Cowan, J.D. (1981). Specificity and
plasticity of retinotectal connections: a computational model.
J. Neurosci. 1, 1369-1387.

\noindent  Wolpert, L. (1969). Positional information and spatial
pattern of cellular differentiation. J. Theor. Biol. 25, 1-47.

\noindent  Wolpert, L. (1996). One hundred years of positional
information. Trends Genet. 12, 359-364.

\noindent  Yates, P.A., Roskies, A.L., McLaughlin, T., and O'Leary, D.D.M.
(2001). Topographic-specific axon branching controlled by ephrin-As is
the critical event in retinotectal map development. J. Neurosci. 21, 
8548-8563.

\noindent  Yates, P.A., Holub, A.D., McLaughlin, T., Sejnowski, T.J.,
and O'Leary, D.D.M. (2004). Computational modeling of retinotopic map
development to define contributions of EphA-EphrinA gradients, axon-axon
interactions, and patterned activity. J. Neurobiol. 59, 95-113.

\newpage

{\bf \large Figure Captions}

Fig. 1\\
Schematic diagram depicting the pathway influencing the axon 
guidance molecules A, B, and C. The transcription factors Emx2 
and Pax6 repress each
other with strengths $v_{1}$ and $v_{2}$. Emx2 represses FGF8 with
strength $w_{1}$, and FGF8 represses Emx2 with strength $w_{2}$. 
The renormalized expression concentration $f$ of FGF8 or some other
signal activated by it serves as
a morphogen for the guiding molecules A, B, and C.

\vspace{0.3cm}

Fig. 2\\
Stationary spatial profiles of the signal $f$ and the guiding molecules
under normal conditions (wild type). (A) Renormalized concentration $f(x)$ 
of FGF8. (B) Concentrations of the guiding molecules A, B, and C. Dashed
line represents the concentration of A, solid line corresponds to B, and 
dashed-dotted line represents the molecule C.
Parameters used: $A_{emx}= 1.34$, $A_{pax}= 1.4$, $A_{fgf}= 0.9$,
$\zeta_{emx}=25.6$, $\zeta_{pax}=27.3$, $\zeta_{fgf}=26.4$, 
$w_{1}=2.4$, $w_{2}=2.1$, $v_{1}=2.6$, $v_{2}=2.7$, 
$\gamma_{A1}= 1.6$, $\gamma_{A2}= -0.4$, $\gamma_{A3}= -2.21$,
$\gamma_{A4}= -2.1$, $\gamma_{A5}= -2.45$, 
$\gamma_{B1}= -0.6$, $\gamma_{B2}= -0.5$, $\gamma_{B3}= 0.4$, 
$\gamma_{B4}= -0.5$, $\gamma_{B5}= -1.0$, 
$\gamma_{C1}= -2.9$, $\gamma_{C2}= -2.5$, $\gamma_{C3}= -2.23$, 
$\gamma_{C4}= -0.6$, $\gamma_{C5}= 1.7$, 
$\theta_{1}= 0.77$, $\theta_{2}= 0.5$,
$\theta_{3}= 0.39$, $\theta_{4}= 0.08$, $\kappa_{A}=0.58$,
$\kappa_{B}=0.9$, $\kappa_{C}=0.55$, $\sigma_{A}= \sigma_{B}= \sigma_{C}=
0.2$. Parameters for the dynamics of TC connections: $D= 0.1$, $L= 40$,
grid size $dx= 0.25$, $\alpha_{i}=3.0$, $\beta_{i}=3.0$ for $i=1,2,3$.

\vspace{0.3cm}

Fig. 3\\
Temporal evolution of the pattern of axonal densities around the cortical 
surface under normal conditions. (A) Initial distribution, 
(B) distribution after $t= 6$,
(C) steady-state. Note the emergence of the heterogeneous pattern.
Solid line represents the profile of $a_{1}$, dashed-dotted line
correspond to $a_{2}$, solid line with open circles represents $a_{3}$,
dashed line corresponds to $a_{4}$, and dotted line corresponds to $a_{5}$.
Parameters are the same as in Fig. 2.

\vspace{0.3cm}

Fig. 4\\
Temporal evolution of the pattern of TC connectivity under normal conditions.
(A) Initial distribution with no TC connectivity 
(all lines collapse onto one), (B) distribution after 
$t= 6$, (C) steady-state. Note the emergence of sharp bordered areas with
different axon types. Regions of high values of $c_{1}$, $c_{3}$, and $c_{5}$ 
correspond to areas M1, S1, and V1, respectively. 
Solid line represents the profile of $c_{1}$, dashed-dotted line
correspond to $c_{2}$, solid line with open circles represents $c_{3}$,
dashed line corresponds to $c_{4}$, and dotted line corresponds to $c_{5}$.
(D) Schematic stationary pattern of emerged areas on the cortical surface,
corresponding to the TC connectivity in (C).
Parameters are the same as in Fig. 2.

\vspace{0.3cm}

Fig. 5\\
Stationary spatial profiles of the signal $f$, the axon guidance molecules,
axonal densities, TC connectivity, and cortical area pattern
when the transcription factor Emx2 is not expressed. For Emx2 mutants 
the distributions of signal $f(x)$ (A), guiding molecules (B),
density of axons (C), TC connectivity (D), and cortical areas (E) 
all shift posteriorly. In (A) the dashed line corresponds to the control
distribution of $f(x)$ from fig. 2A. Parameters are the same as in Fig. 2, 
except for $A_{emx}= 0$.

\vspace{0.3cm}

Fig. 6\\
Stationary spatial profiles of the signal $f$, the axon guidance molecules,
axonal densities, TC connectivity, and cortical area pattern
when the transcription factor Pax6 is not expressed. For Pax6 mutants the 
distributions of signal $f(x)$ (A), guiding molecules (B), density of axons 
(C), TC connectivity (D), and cortical areas (E) all shift anteriorly. 
In (A) the dashed line corresponds to the control distribution of $f(x)$ 
from fig. 2A. Parameters are the same as in Fig. 2, except for $A_{pax}= 0$.

\vspace{0.3cm}

Fig. 7\\
The influence of FGF8 on the cortical architecture and guiding molecules. 
(A), (B), and (C) Overexpression of FGF8. (D), (E), and (F) underexpression 
of FGF8. Note the opposite shifts of the areas in these cases (compare
(B), (C) with (E), (F), respectively). 
Line convention is the same as in Figs. 2 and 4.
Parameters used: (A), (B) $A_{fgf}= 1.6$, $\zeta_{fgf}= 32.8$, (C), (D)
$A_{fgf}= 0.6$, $\zeta_{fgf}= 19.0$. Other parameters are the same as 
in Fig. 2.

\vspace{0.3cm}

Fig. 8\\
Generation of two partly symmetric S1 areas when two sources of FGF8
are present. (A) Stationary spatial pattern of TC connectivity and
(B) corresponding area pattern on the cortical surface.
Note the mirror symmetry effect, i.e. not only two S1 areas ($c_{3}$) 
are generated, but
also two M1 fields ($c_{1}$) are present (one regular, second in
the posterior end), and V1 area ($c_{5}$) is located in between
two S1 areas. (C) Spatial profile of the signal $f$. Note a change
of shape and the appearance of a minimum. (D) Stationary concentrations
of the guiding molecules. Note that the molecule B is expressed broadly
and it has two maxima, the molecule A is also partly expressed in the
posterior end, and C is weakly expressed in the central part corresponding
to minimum in $\rho_{B}$. The figures are generated by assuming that
there is a second source of FGF8 at the posterior end with the amplitude
$A_{fgf}'= 1.5$ and the range $\zeta_{fgf}'= 12.0$. Other parameters
are the same as in Fig. 2.

\vspace{0.3cm}

Fig. 9\\
Stationary spatial patterns of TC connectivity in the case of purely
repulsive interactions between guiding molecules and axons. (A)
Pattern of TC connectivity under normal conditions. (B) Pattern of
TC connectivity with Emx2 mutation, i.e. $A_{emx}= 0$. 
(C) Pattern of TC connectivity when
two sources of FGF8 are present similarly as in Fig. 8. Note a mirror
symmetry effect. Parameters used are the same as in Fig. 2 except:
$\gamma_{A1}= -0.08$, $\gamma_{B3}= -0.1$, $\gamma_{C5}= -0.06$.

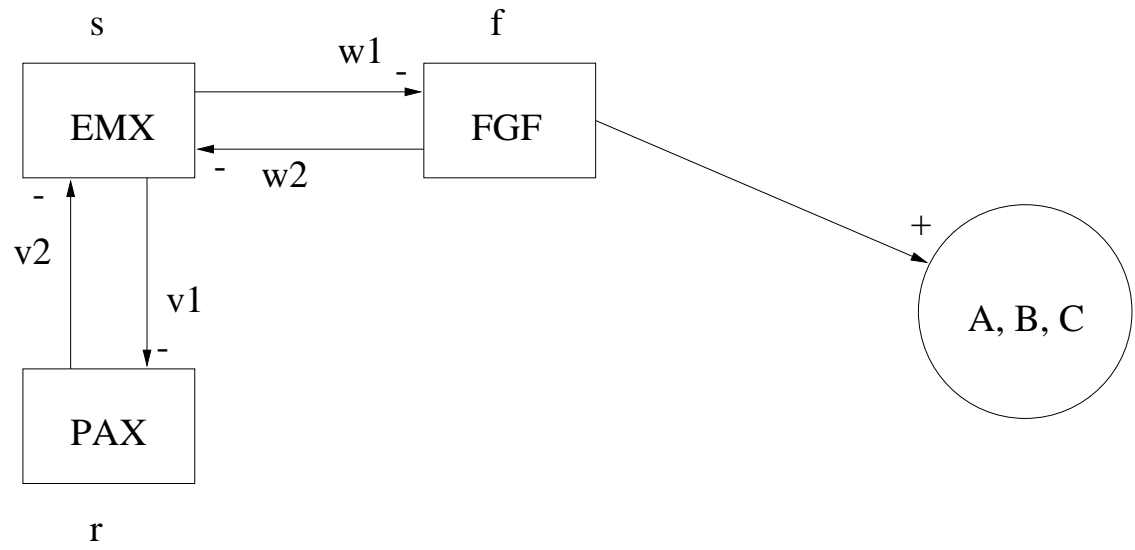
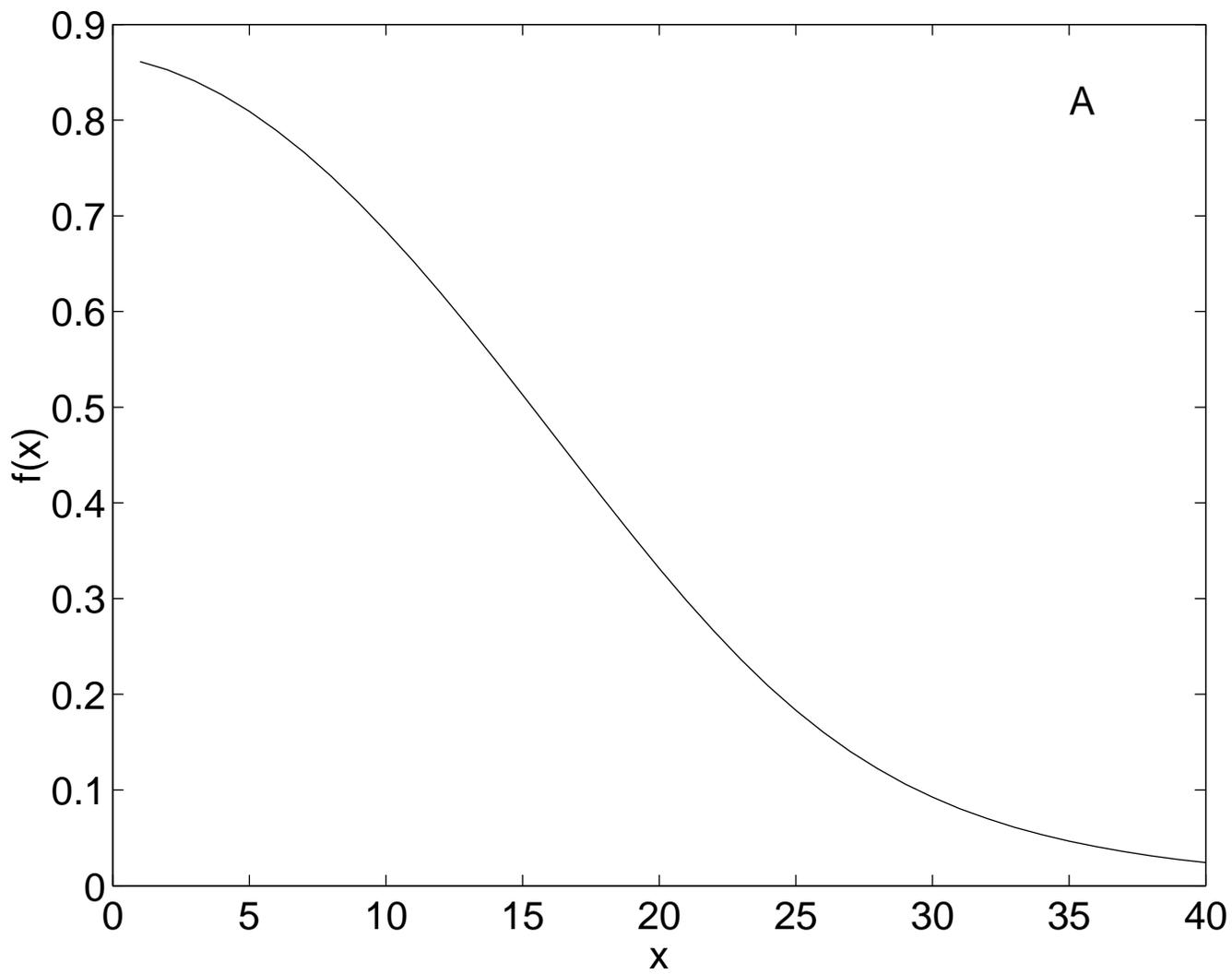
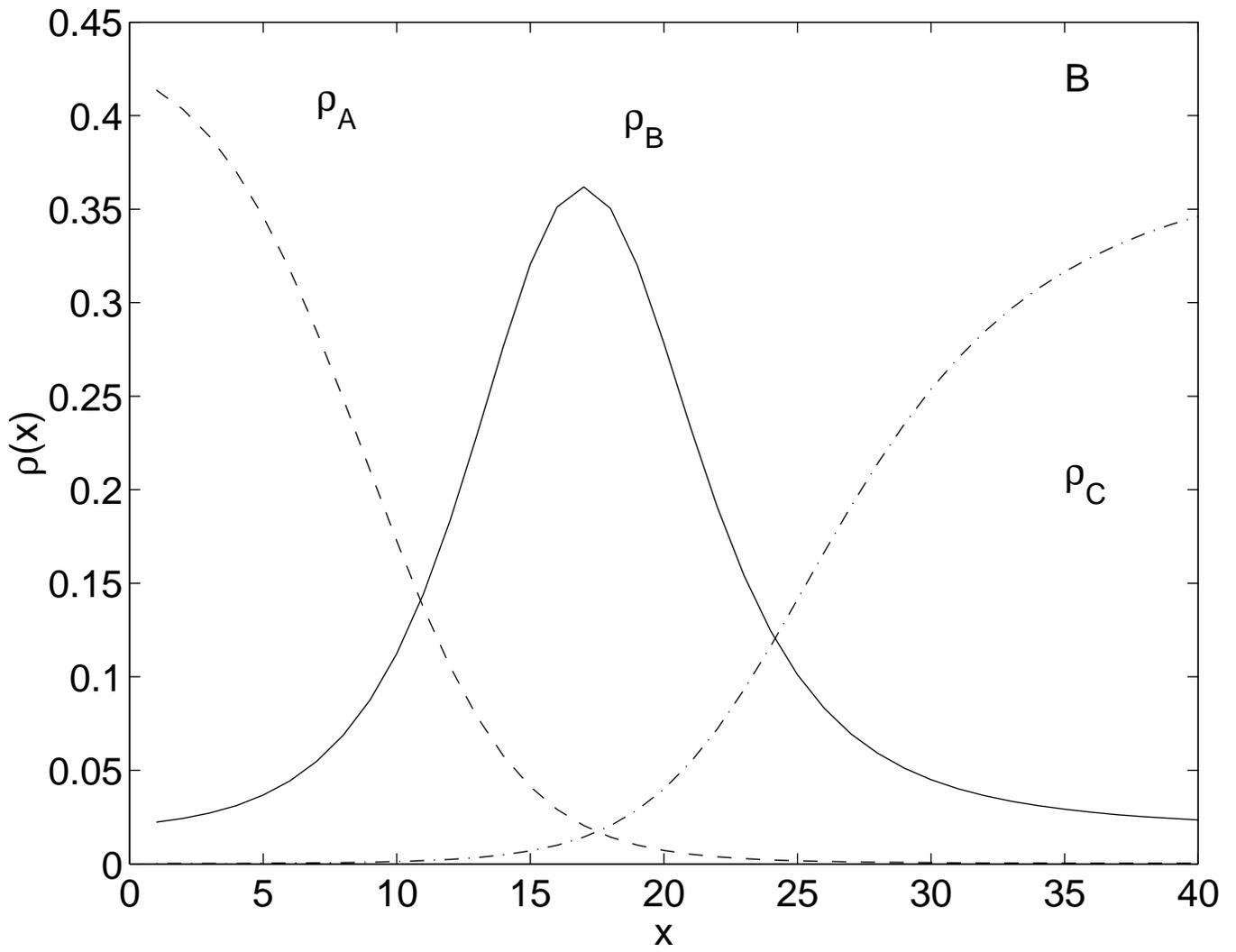
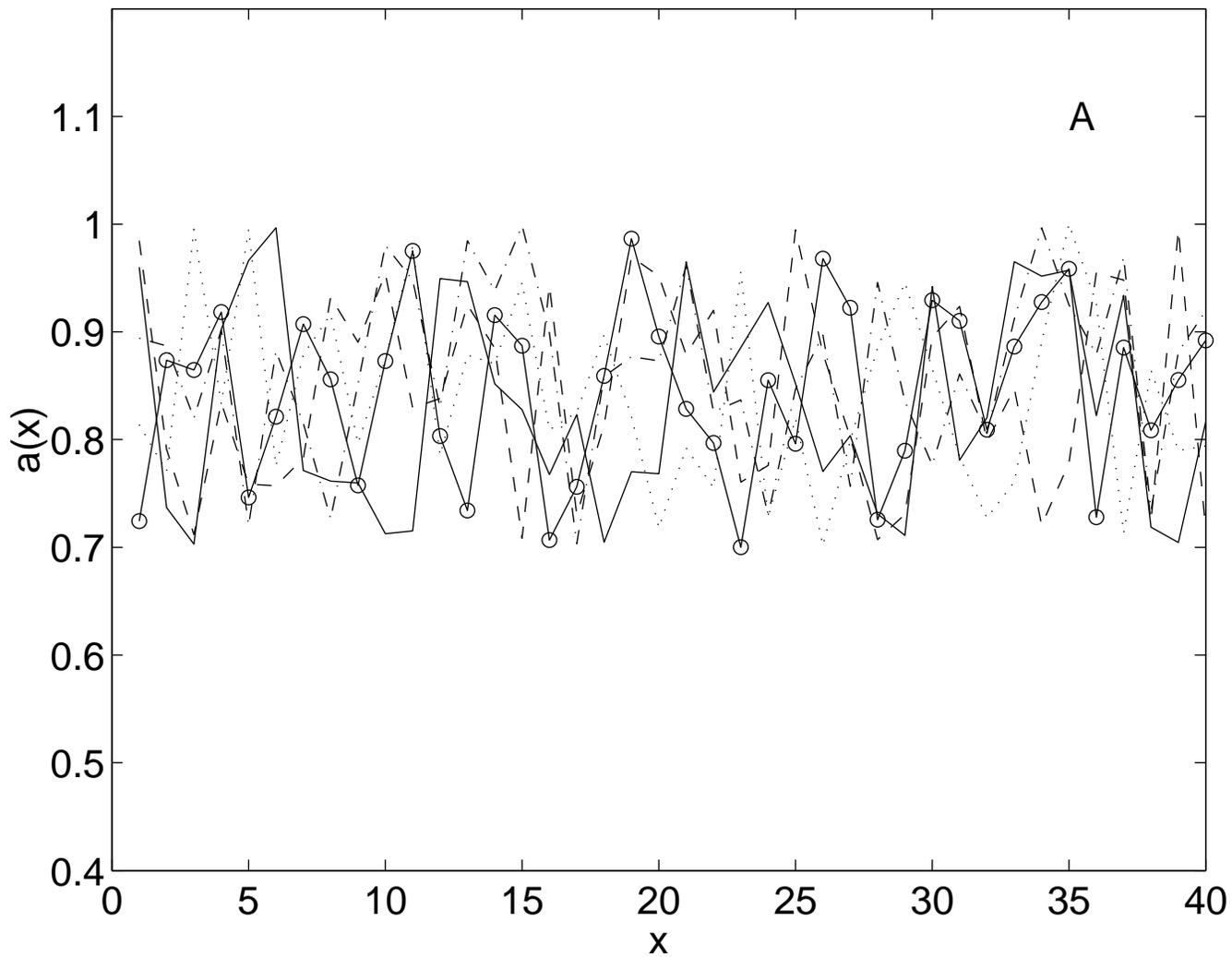
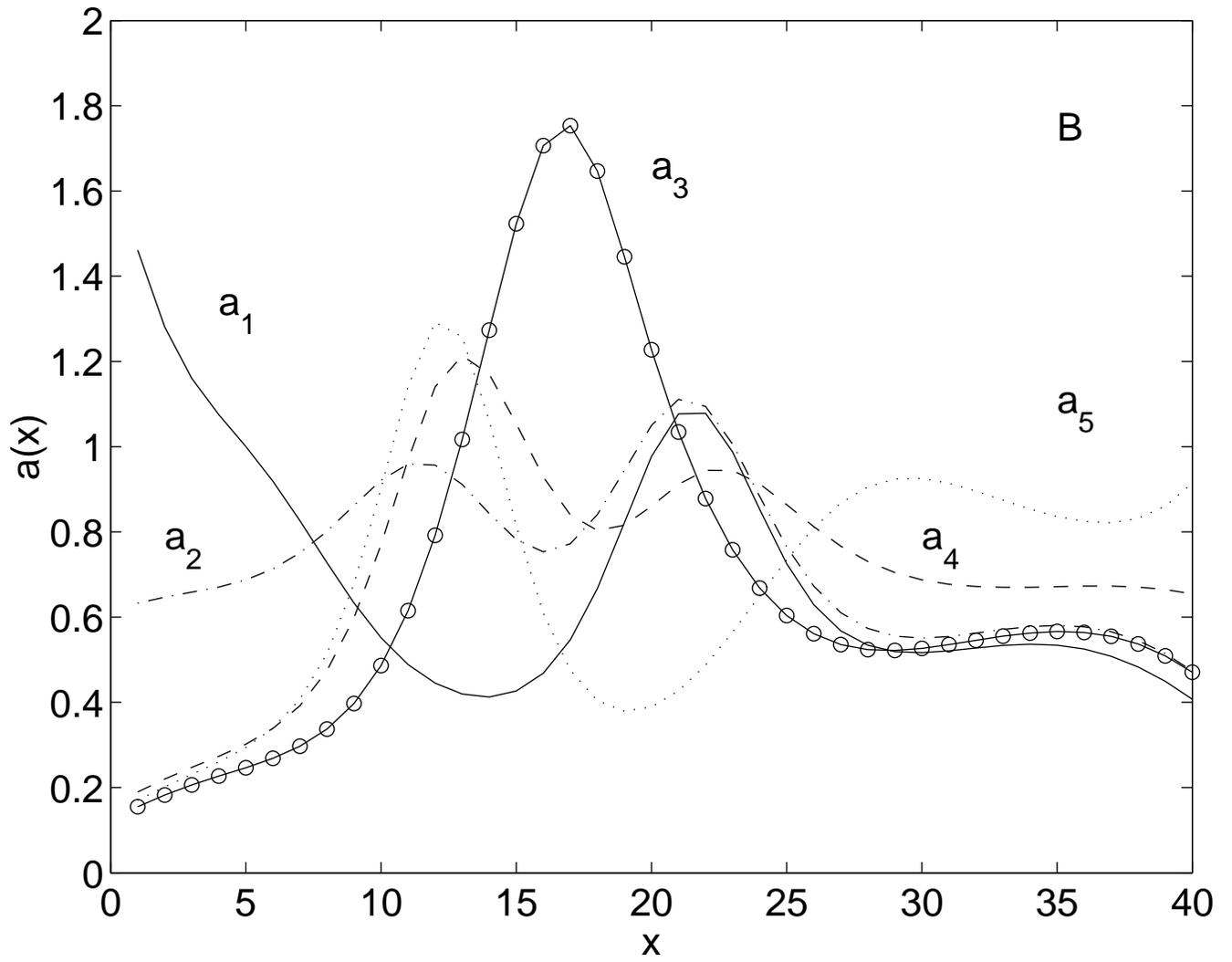
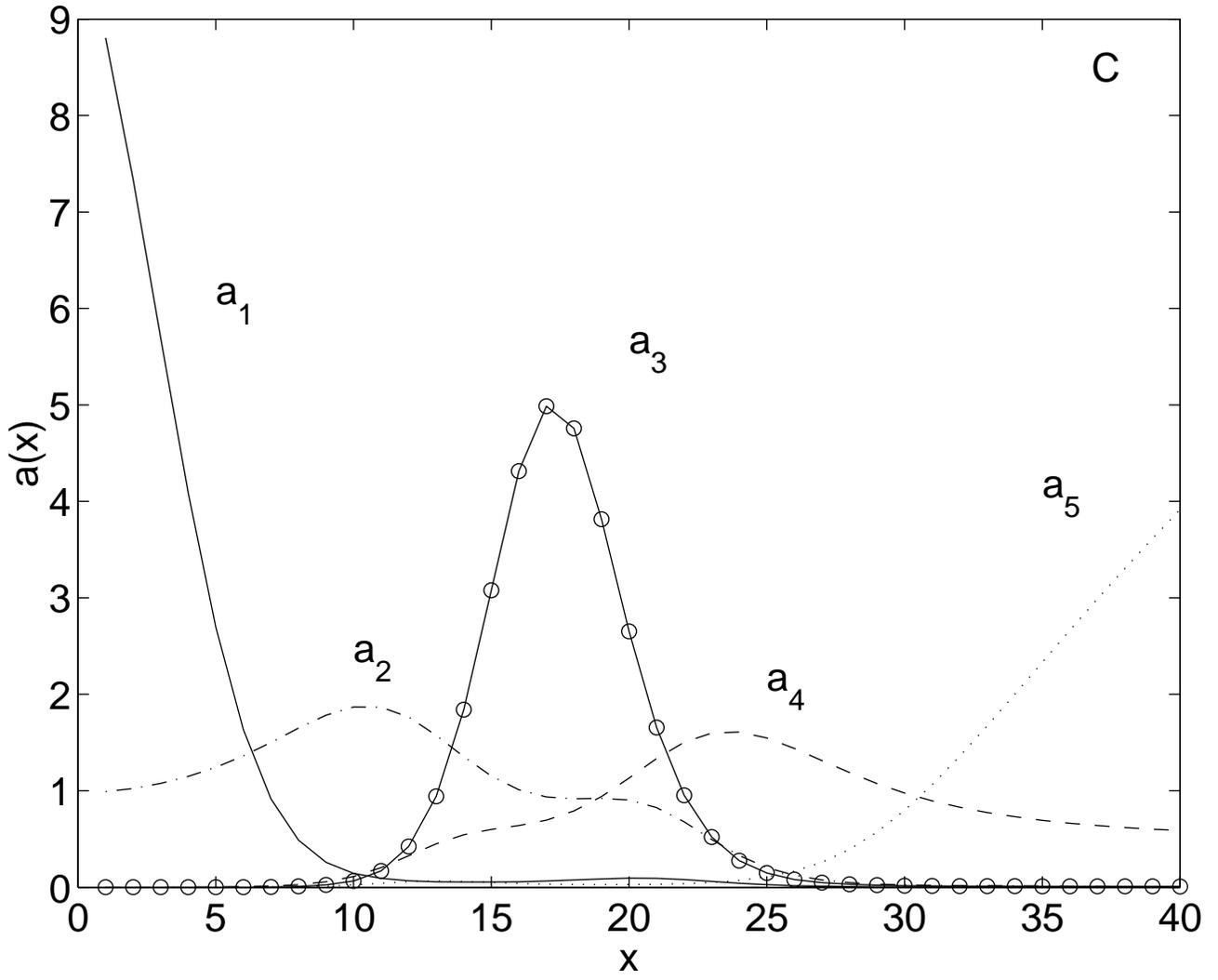
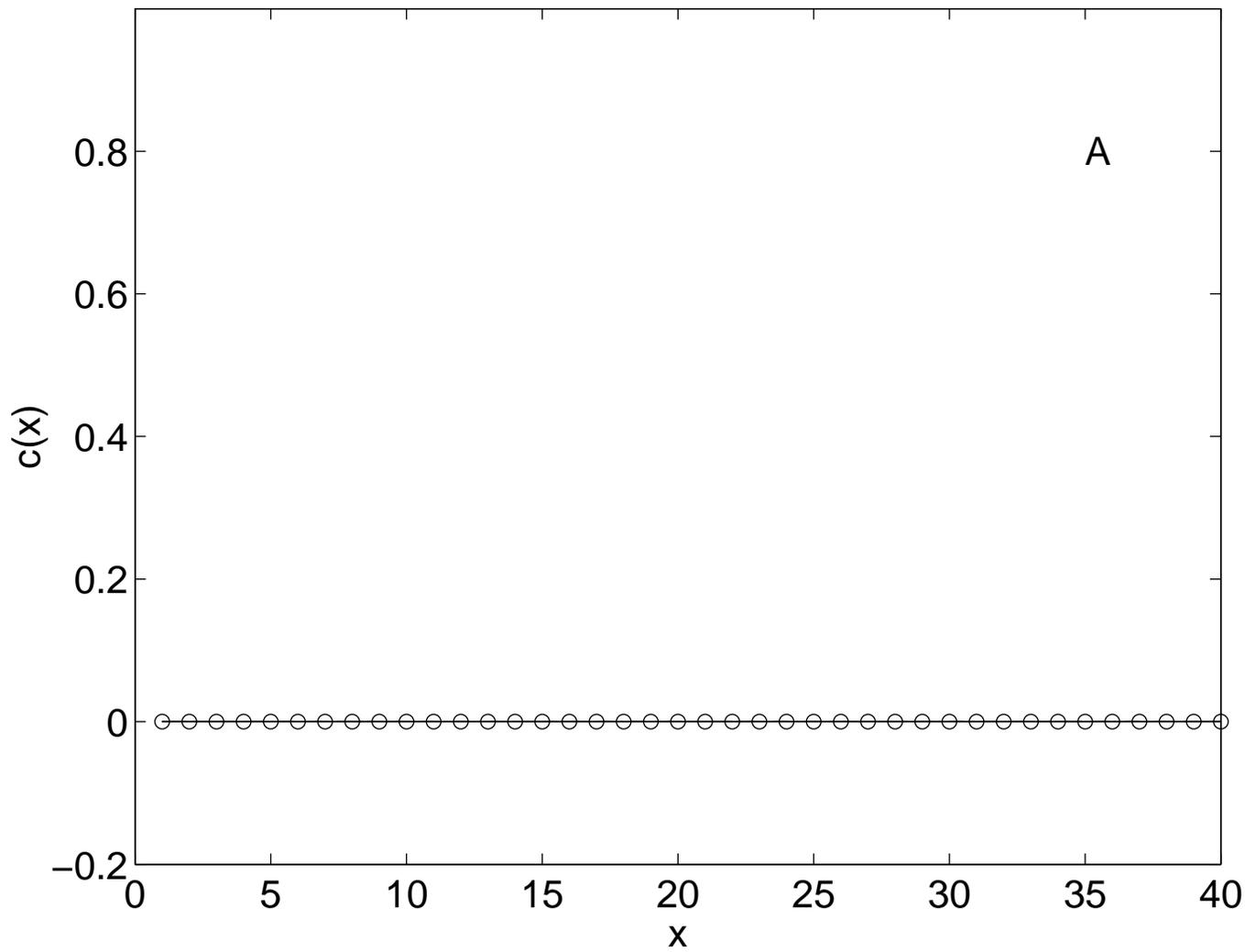
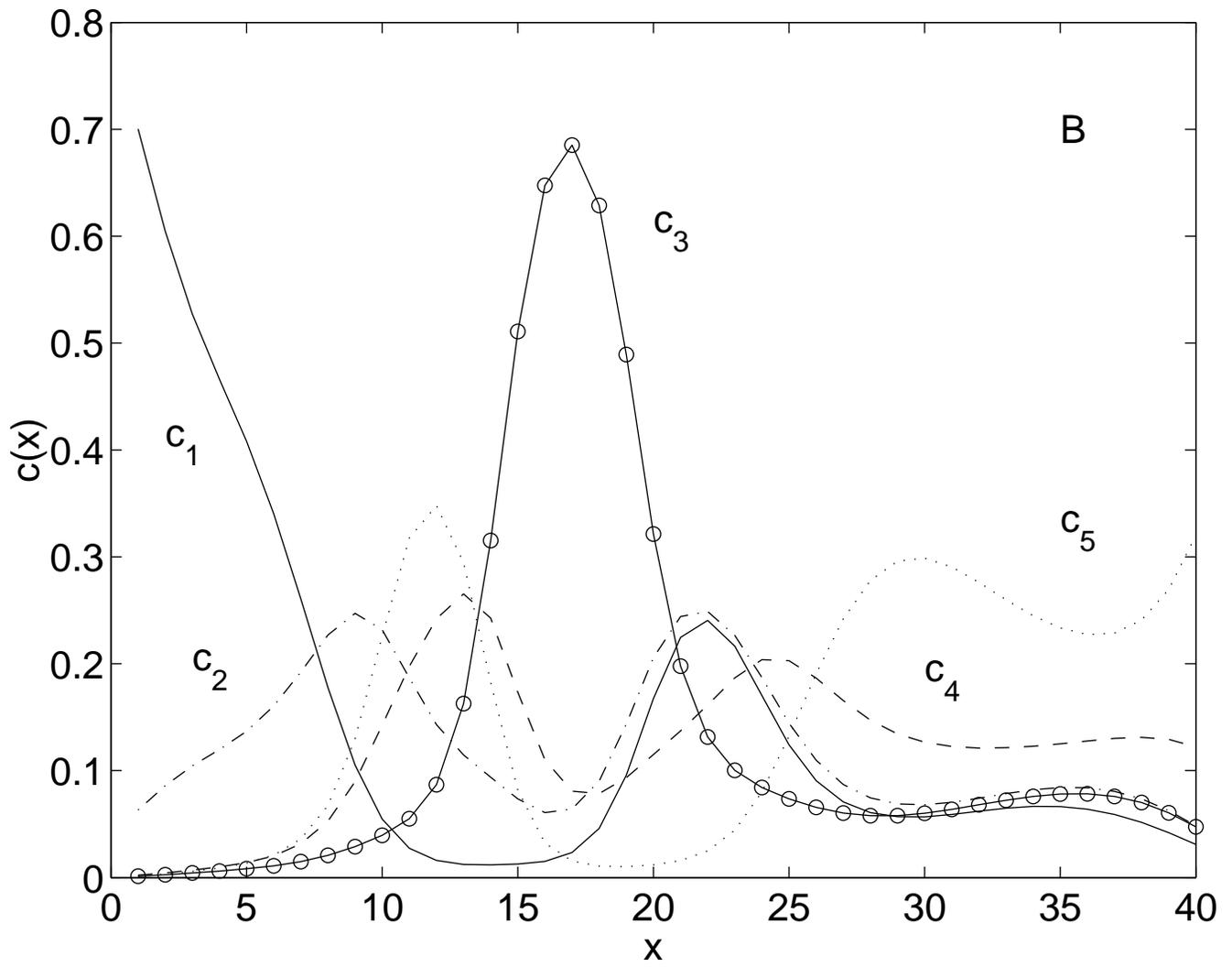
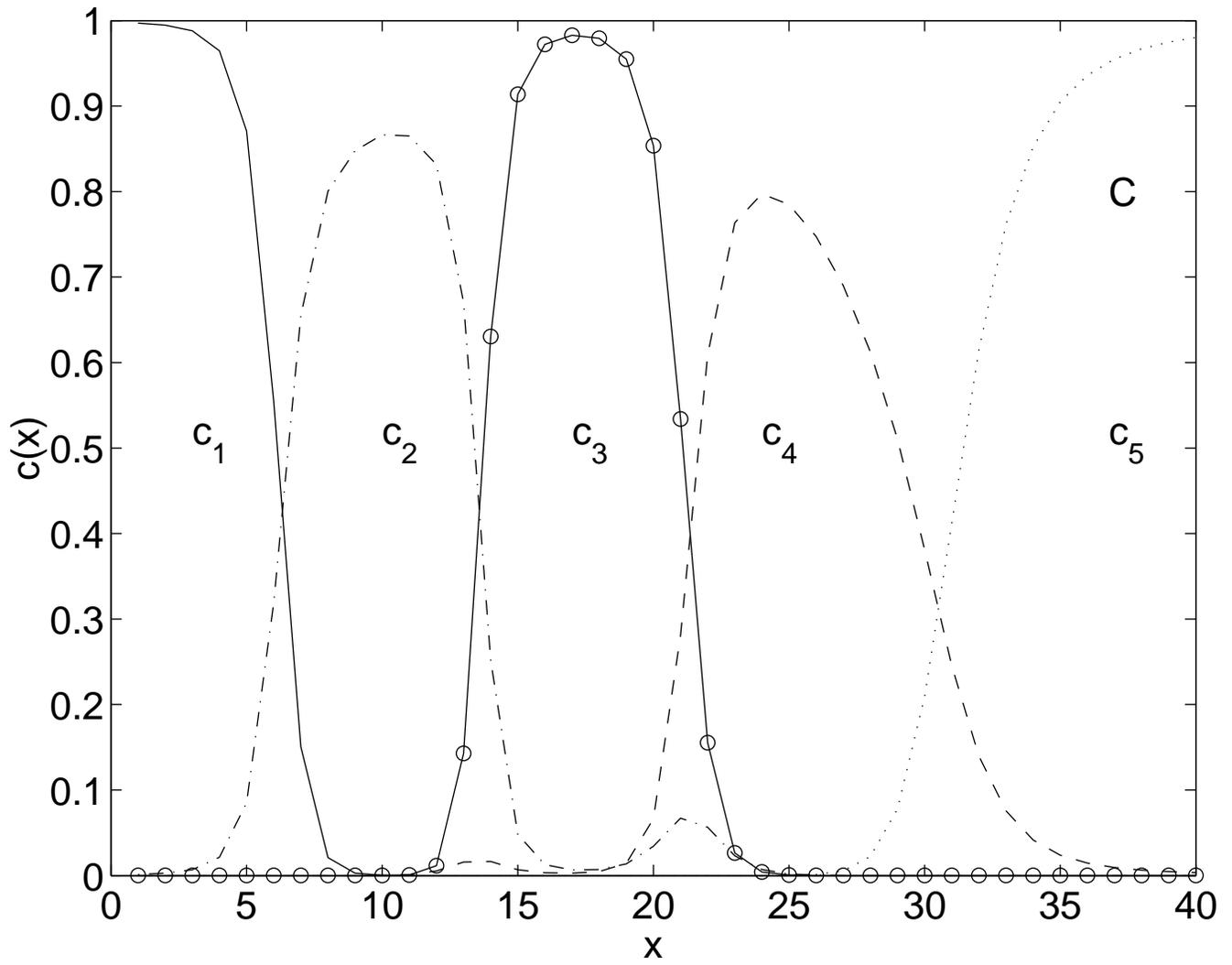
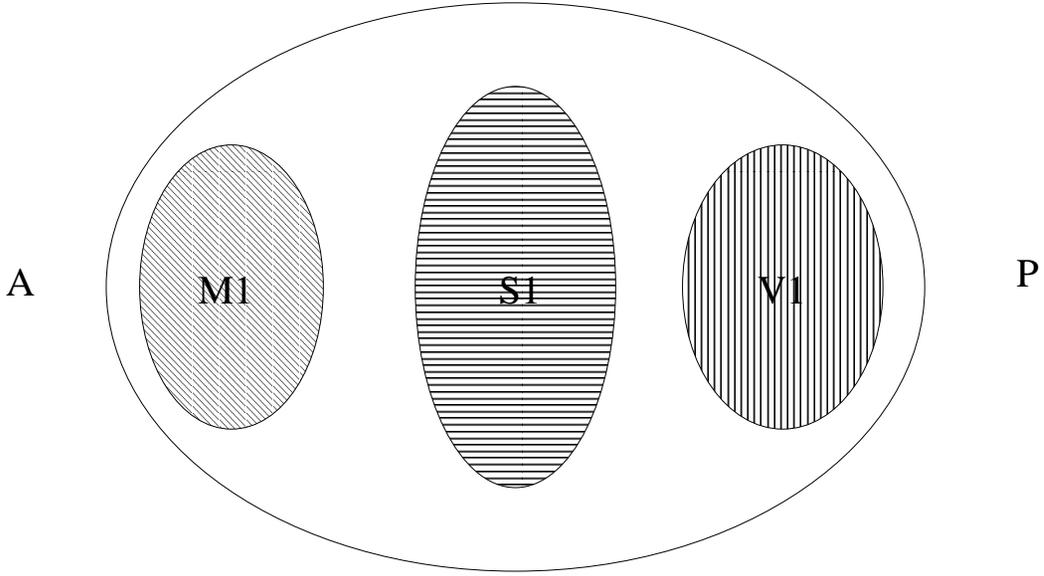
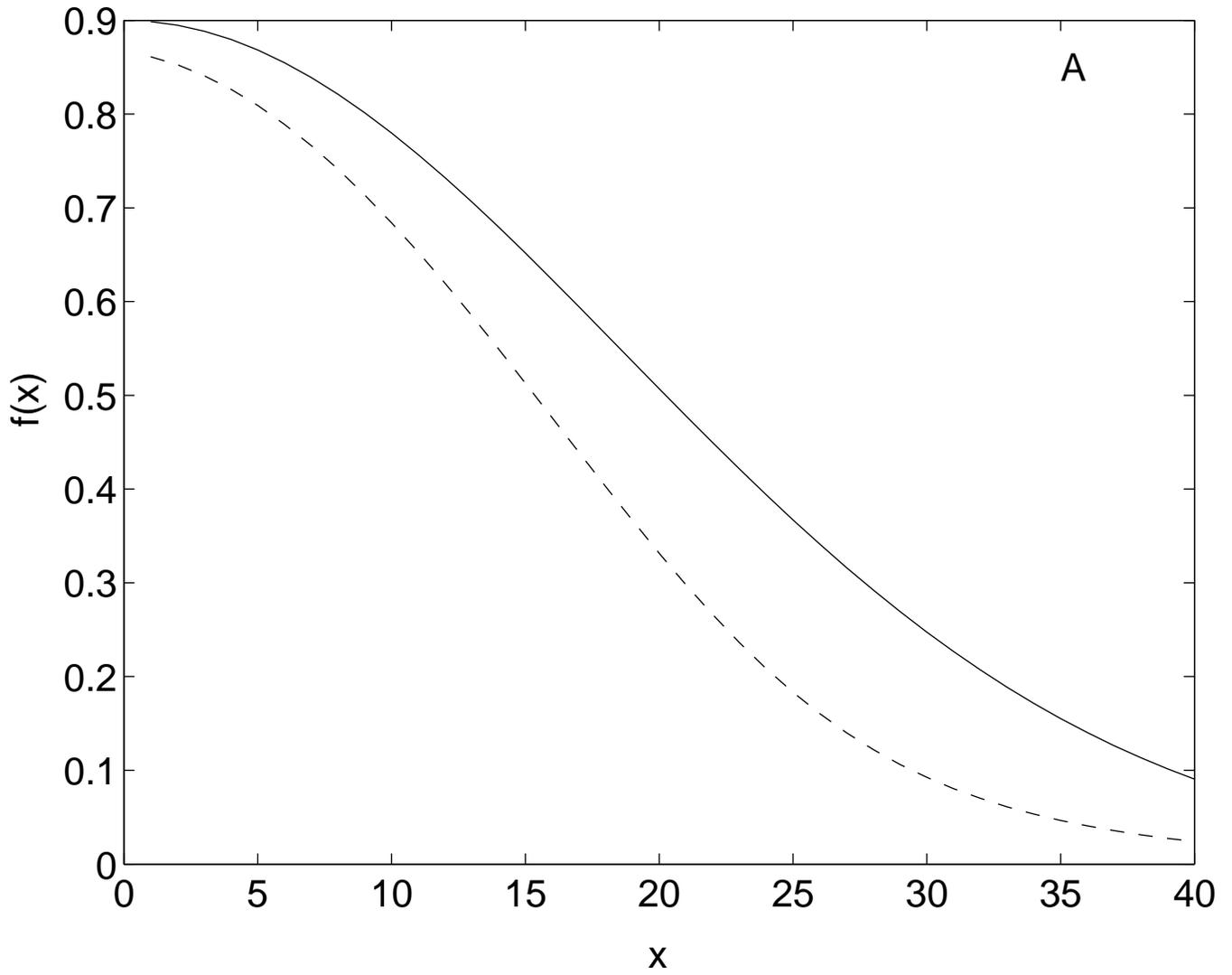
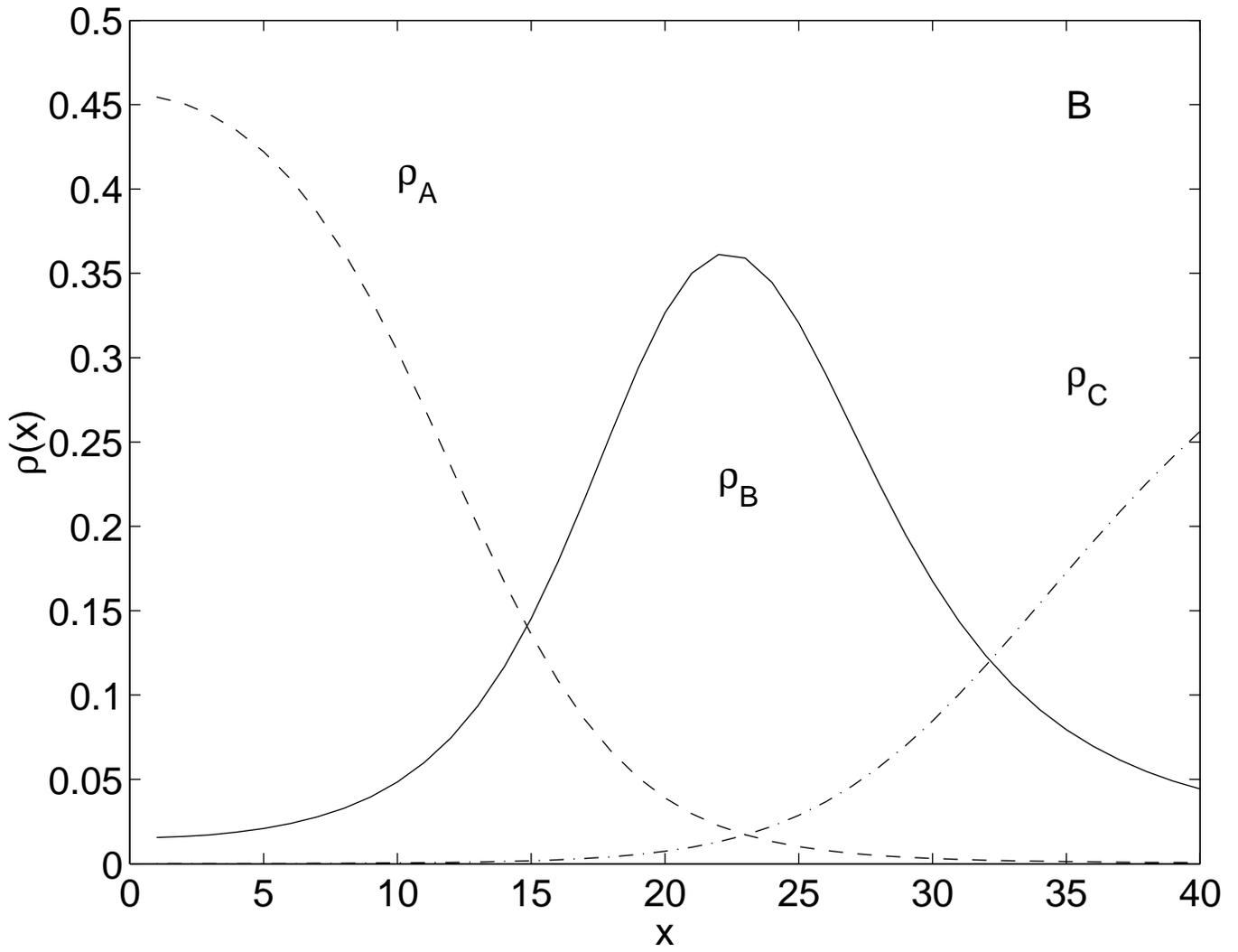
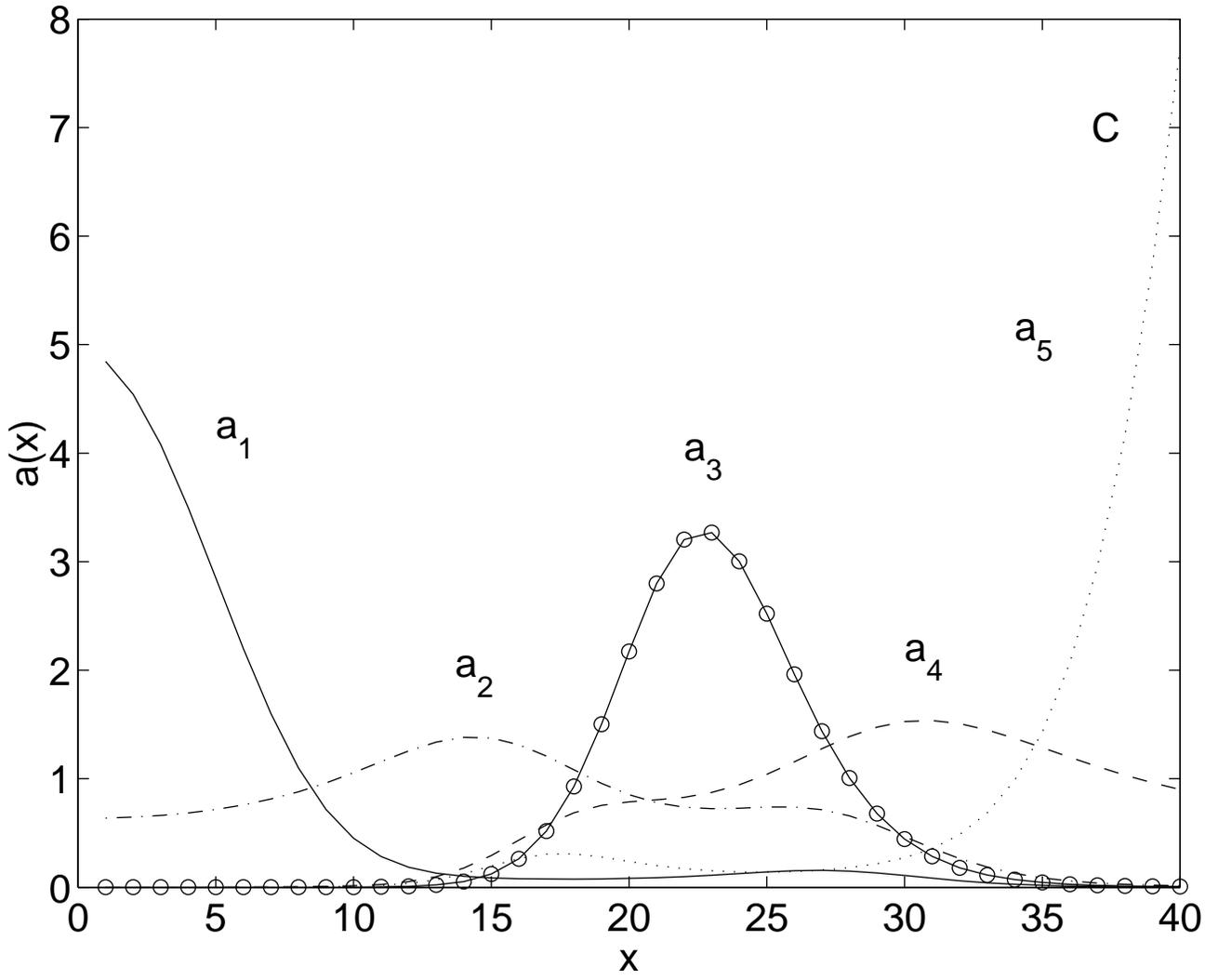
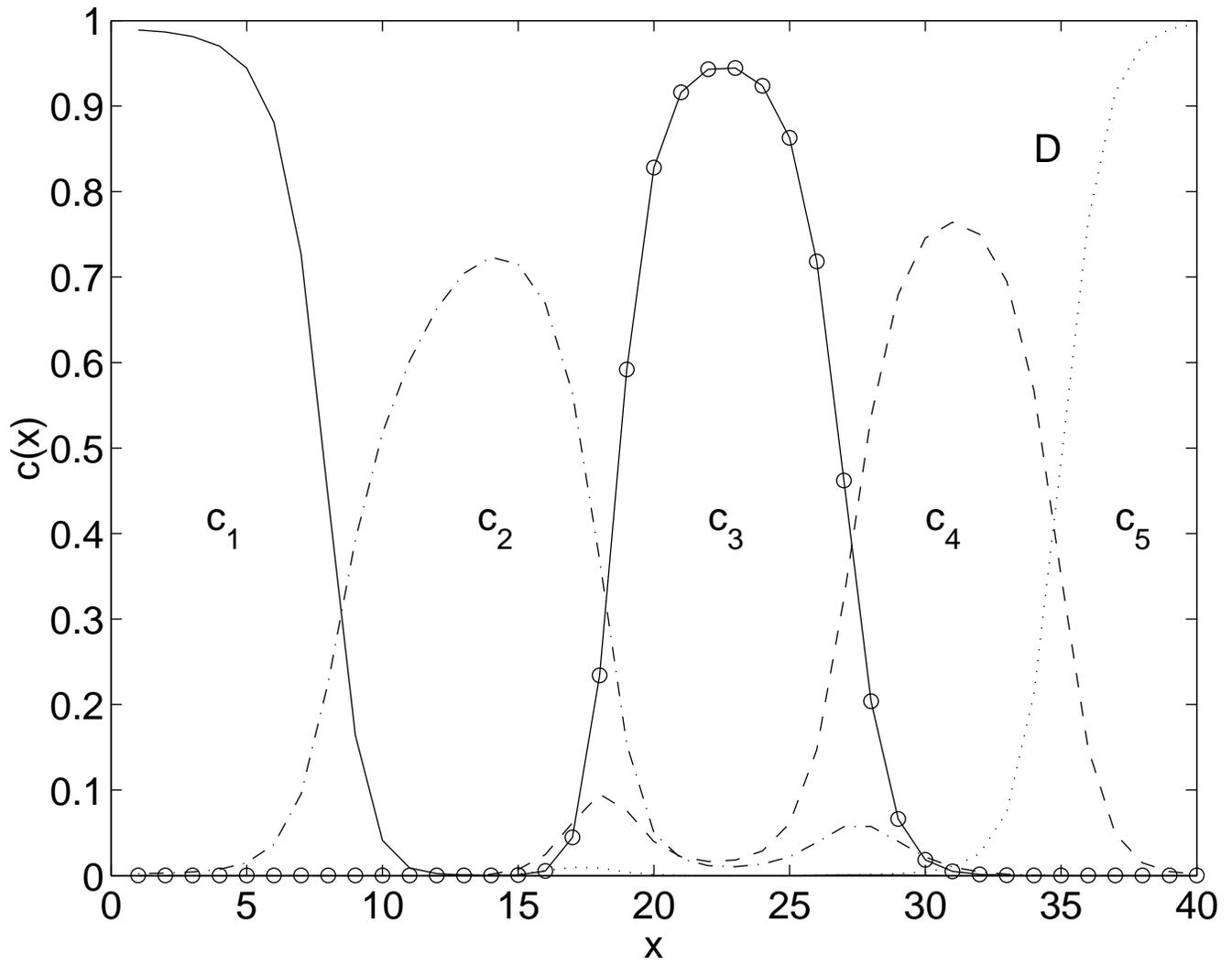
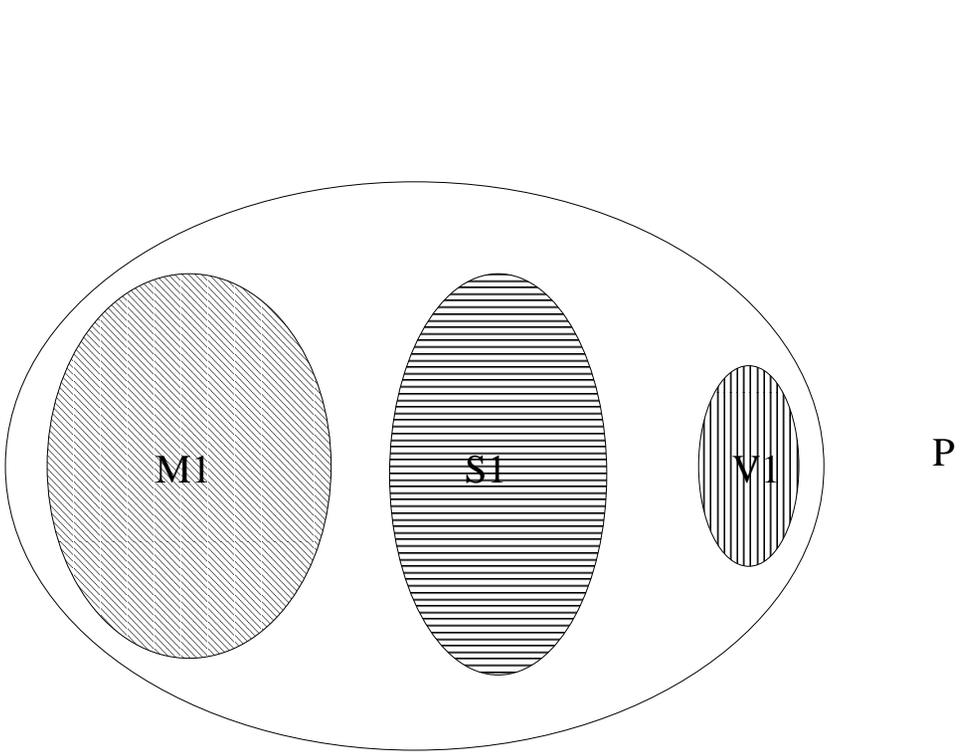
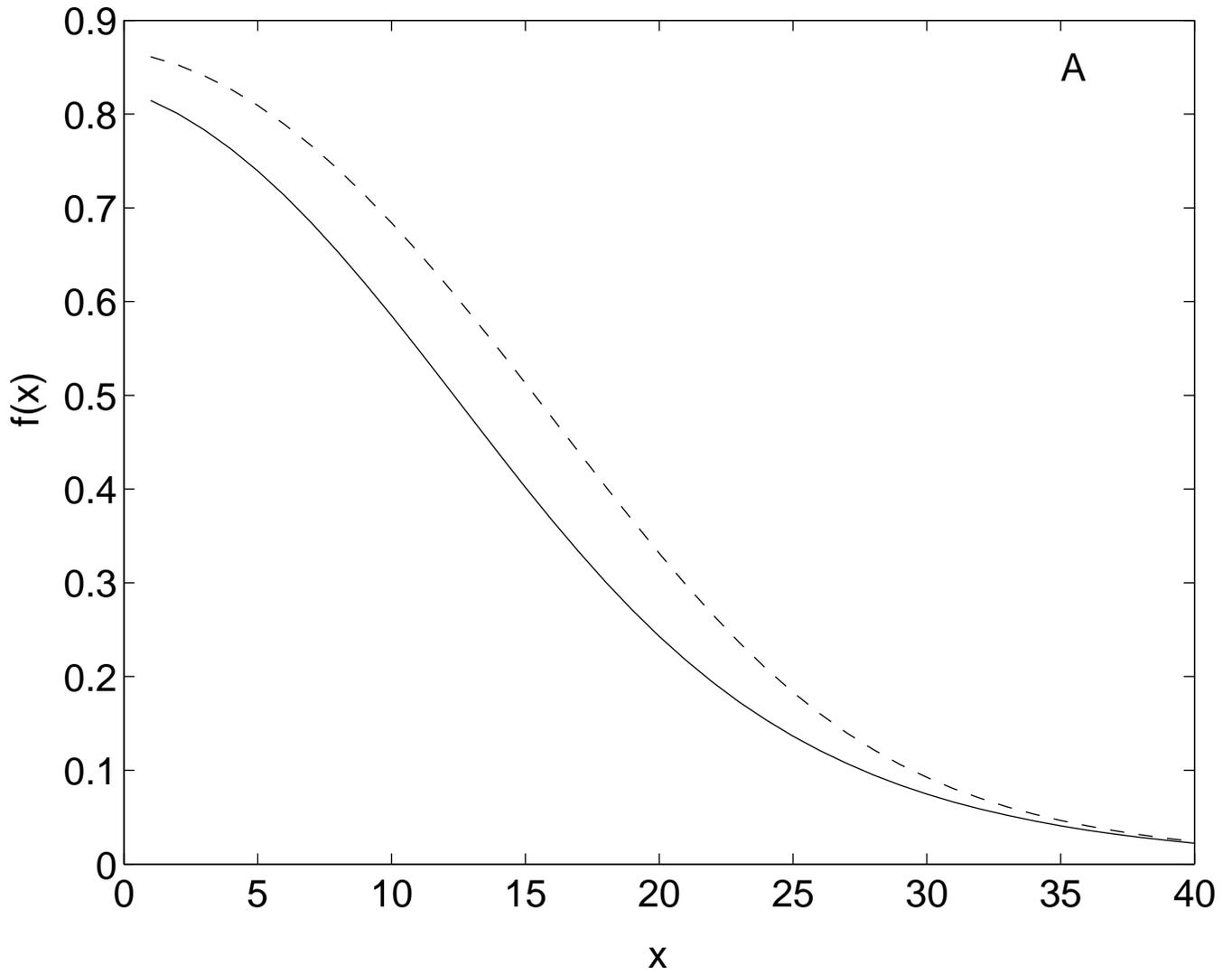
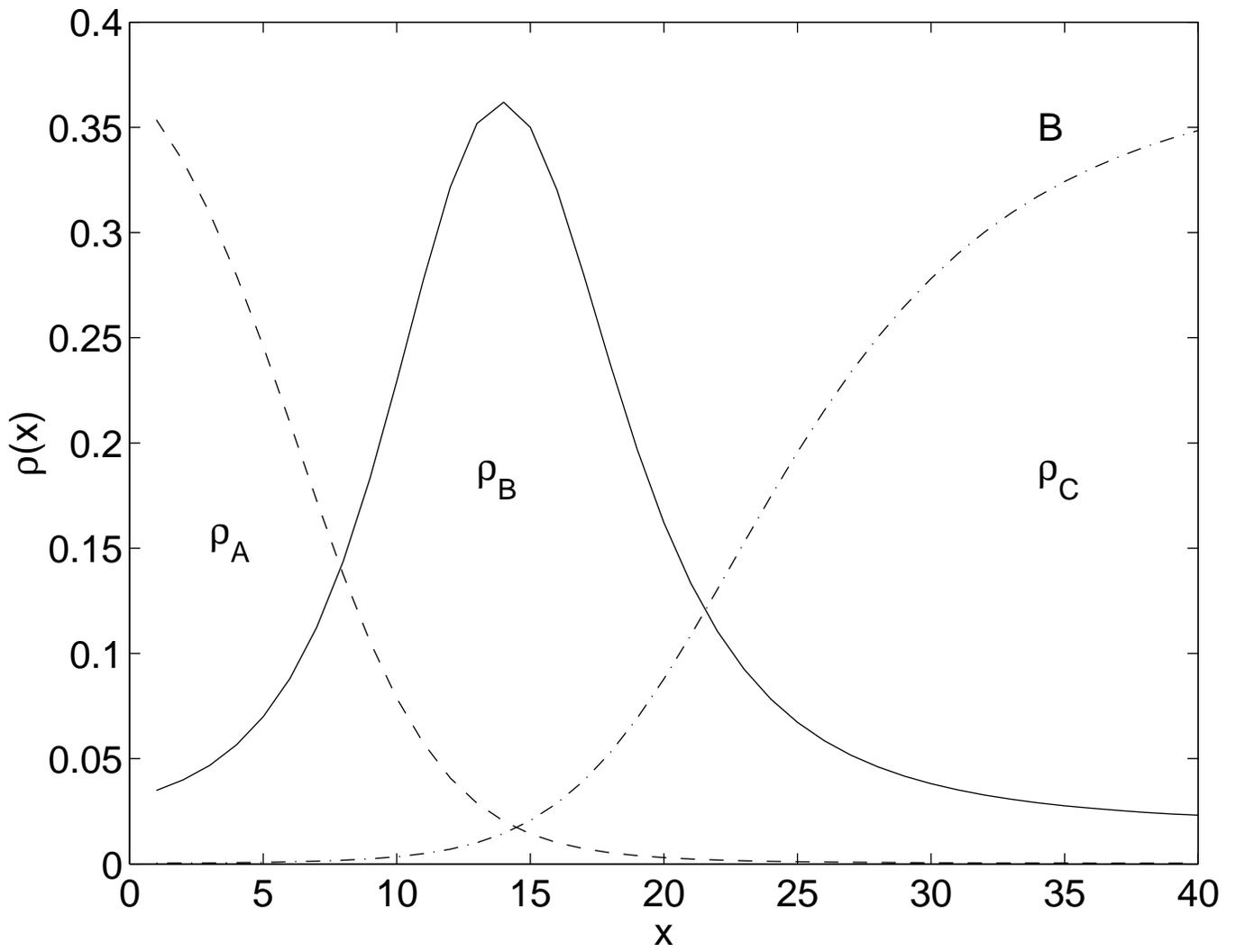
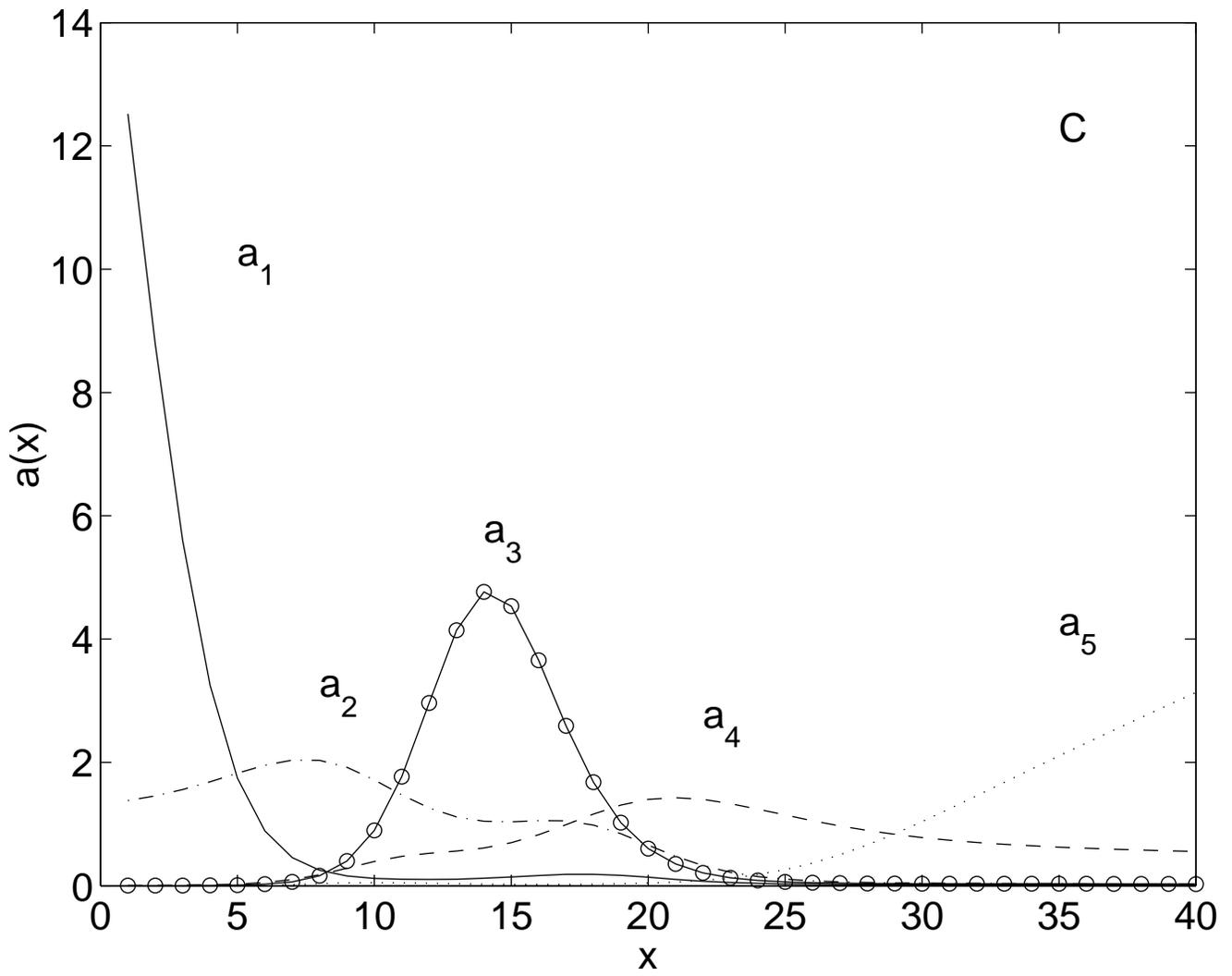
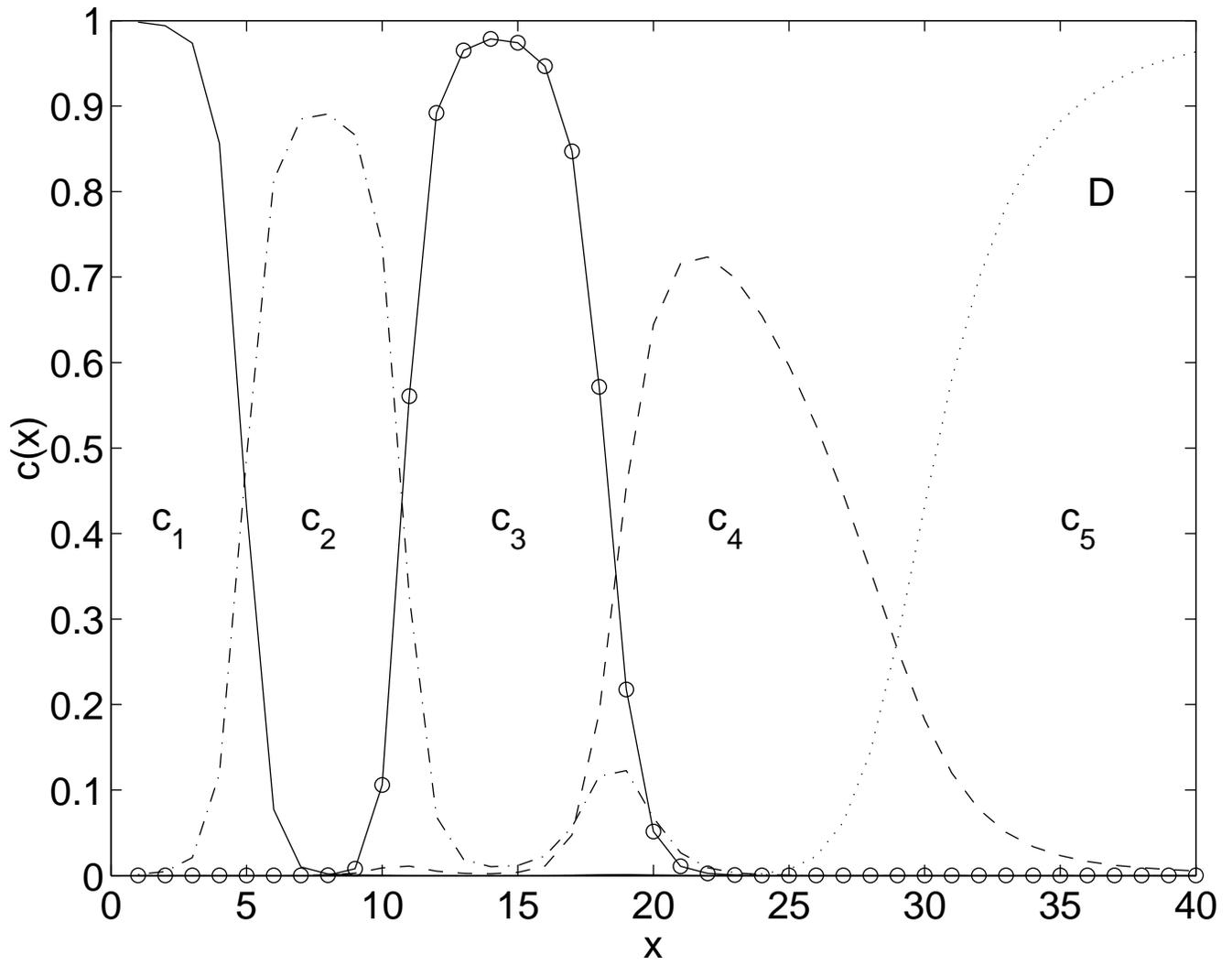
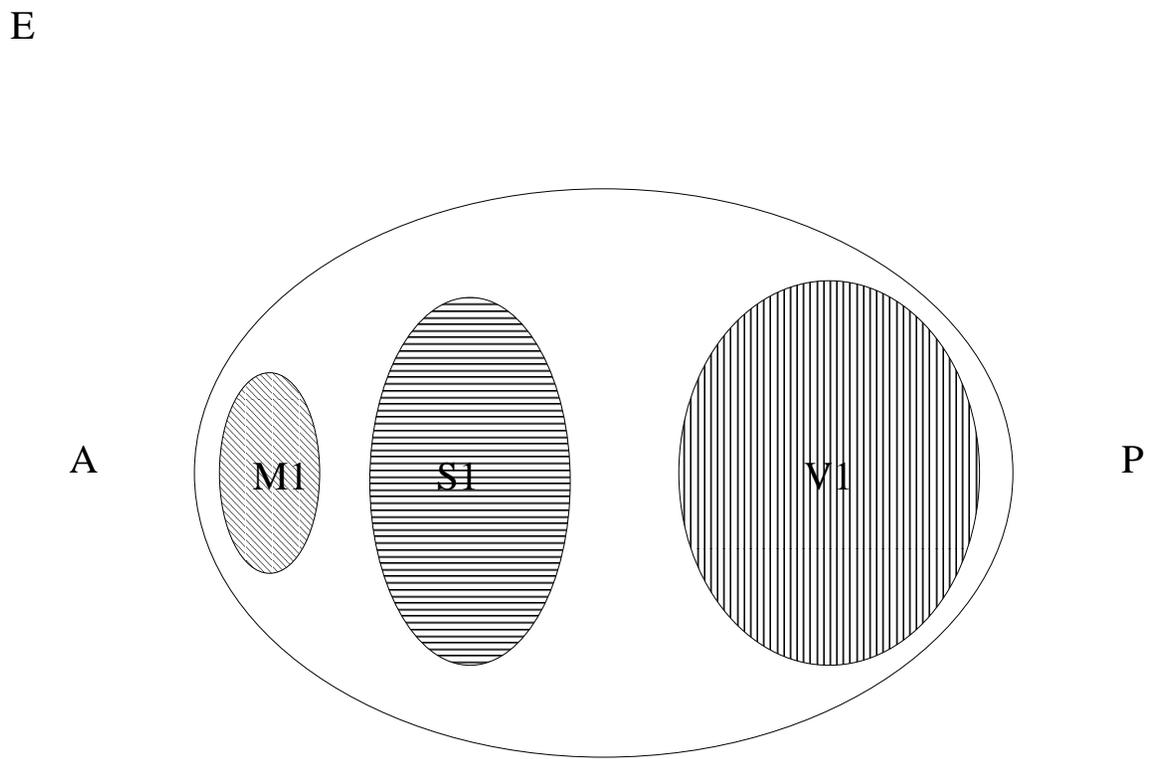
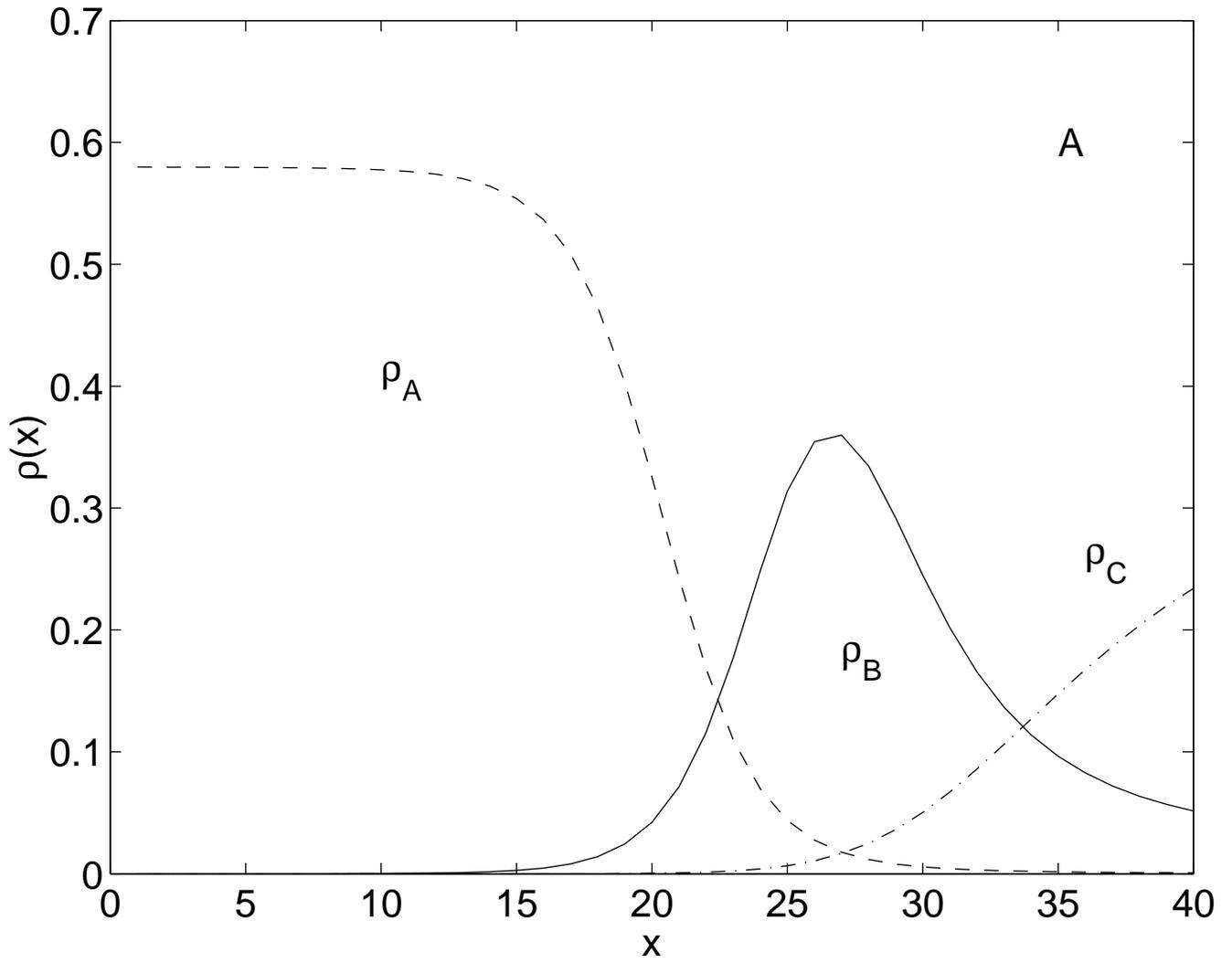
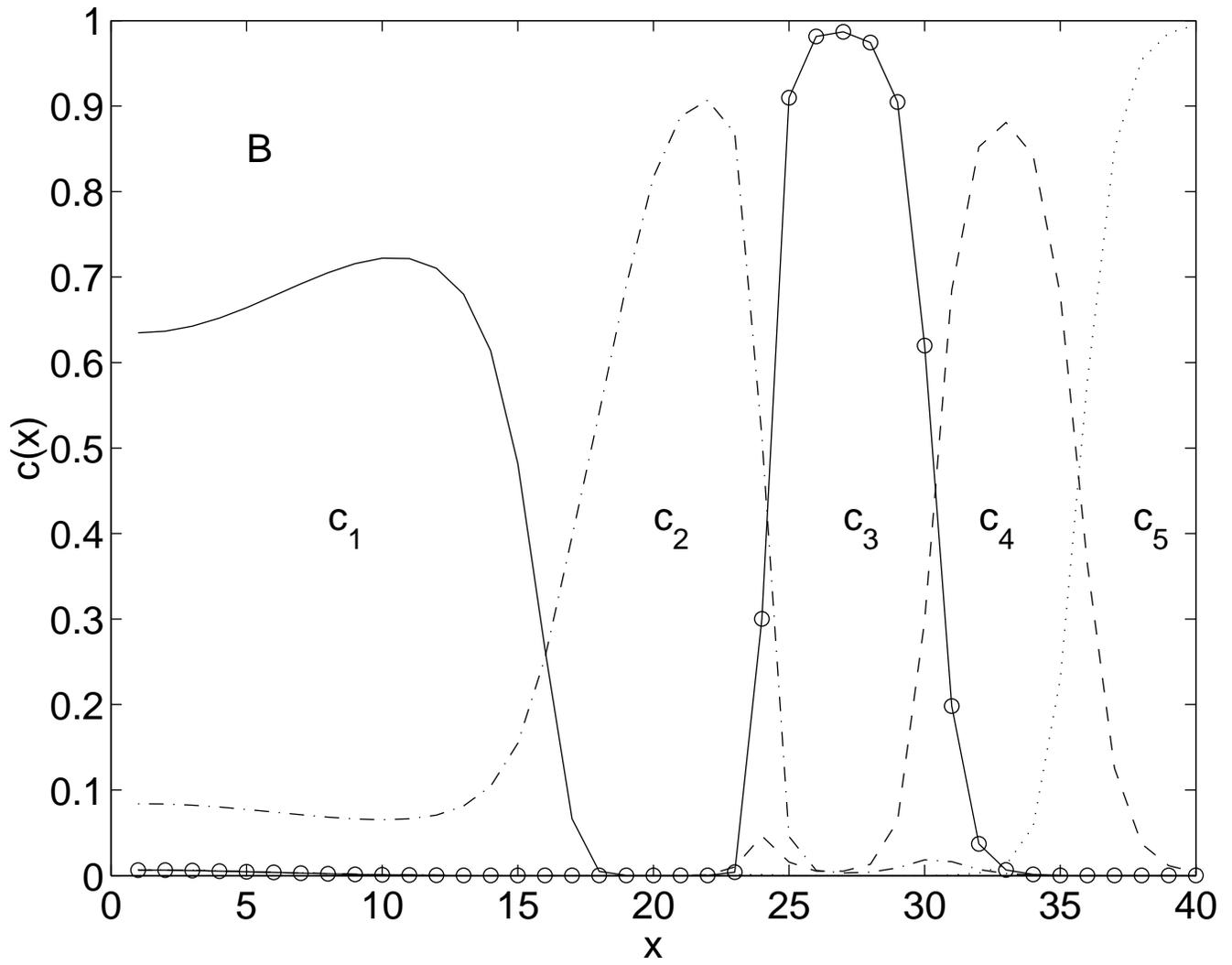
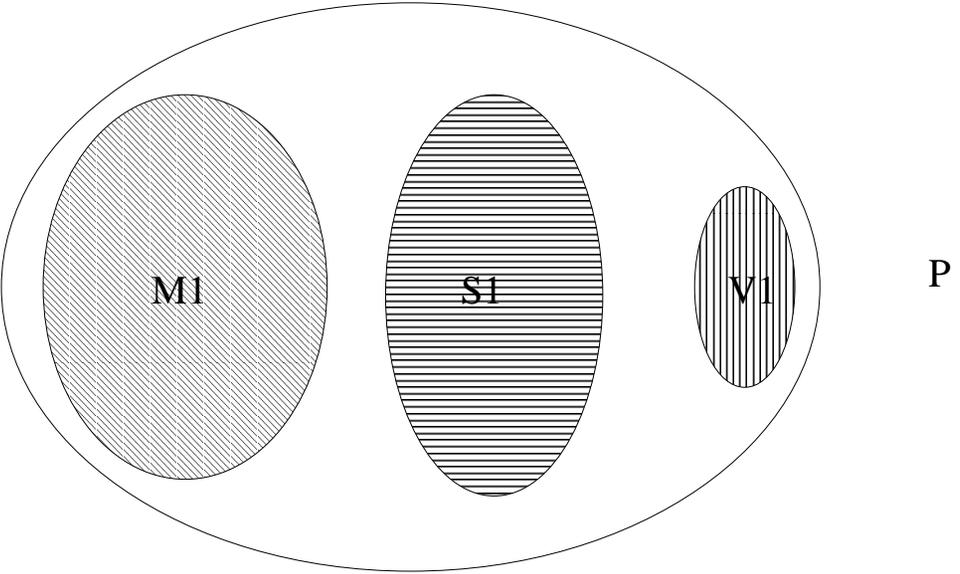
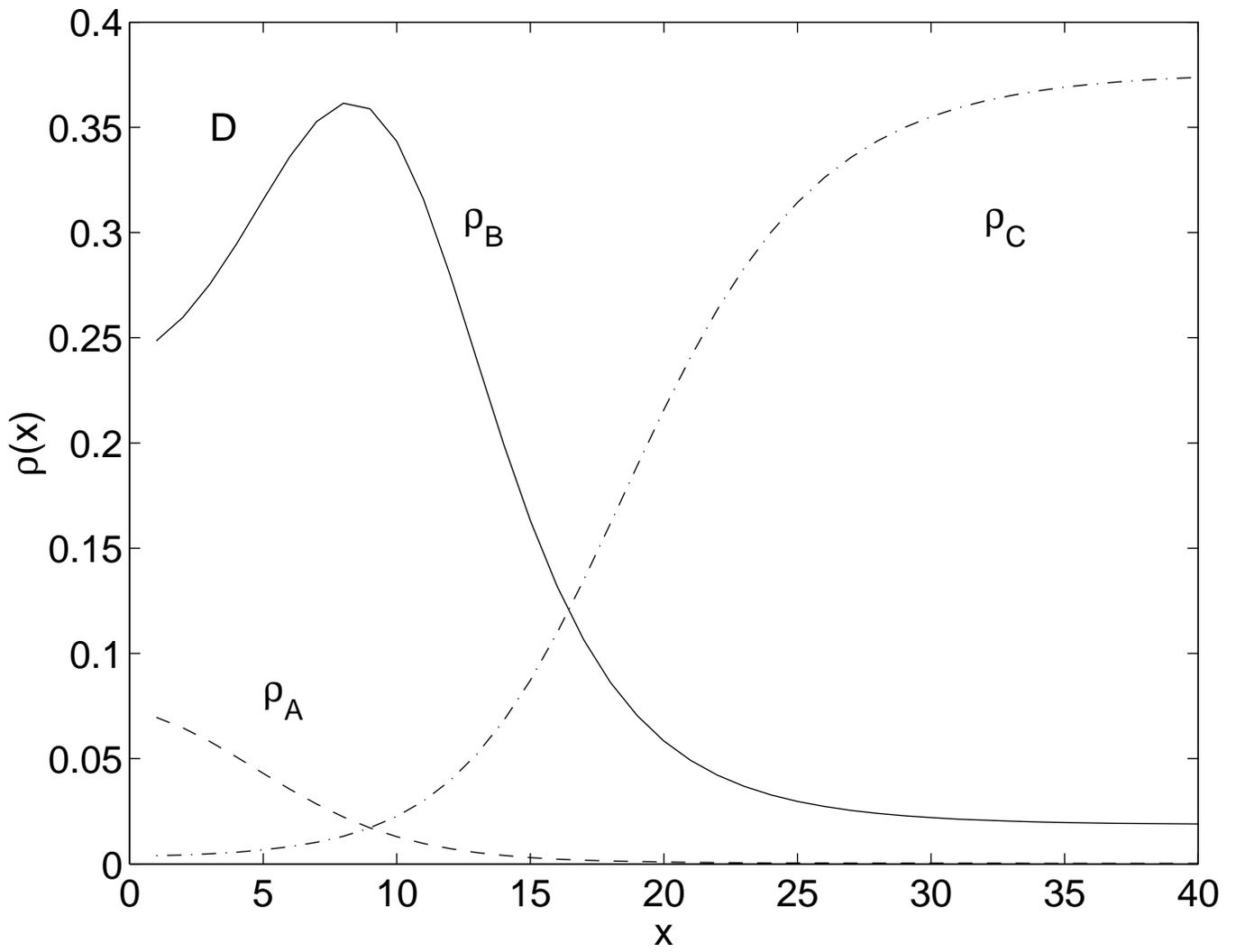
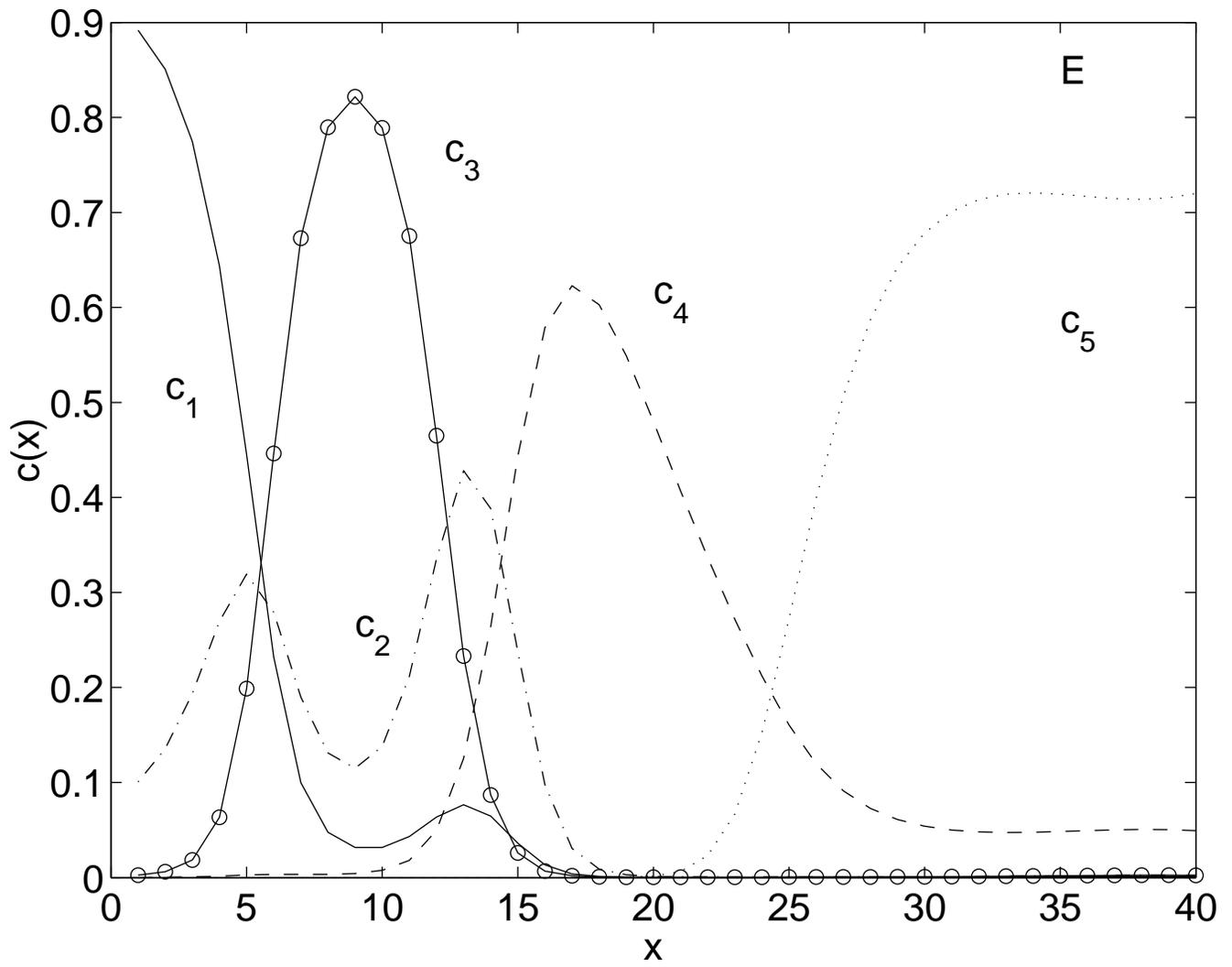
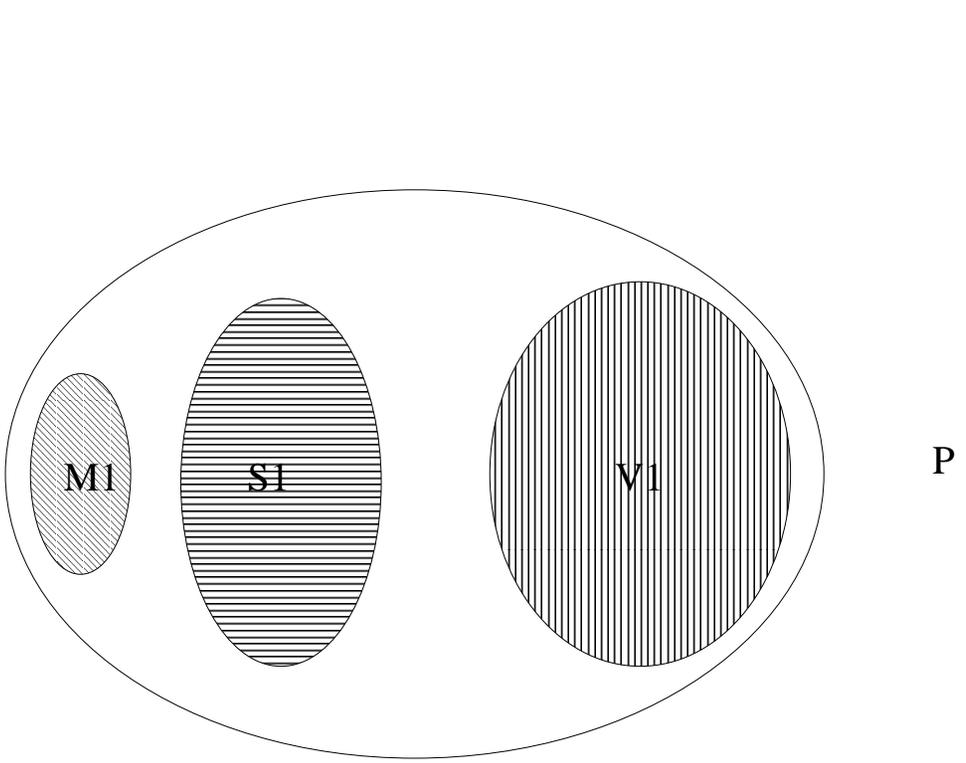
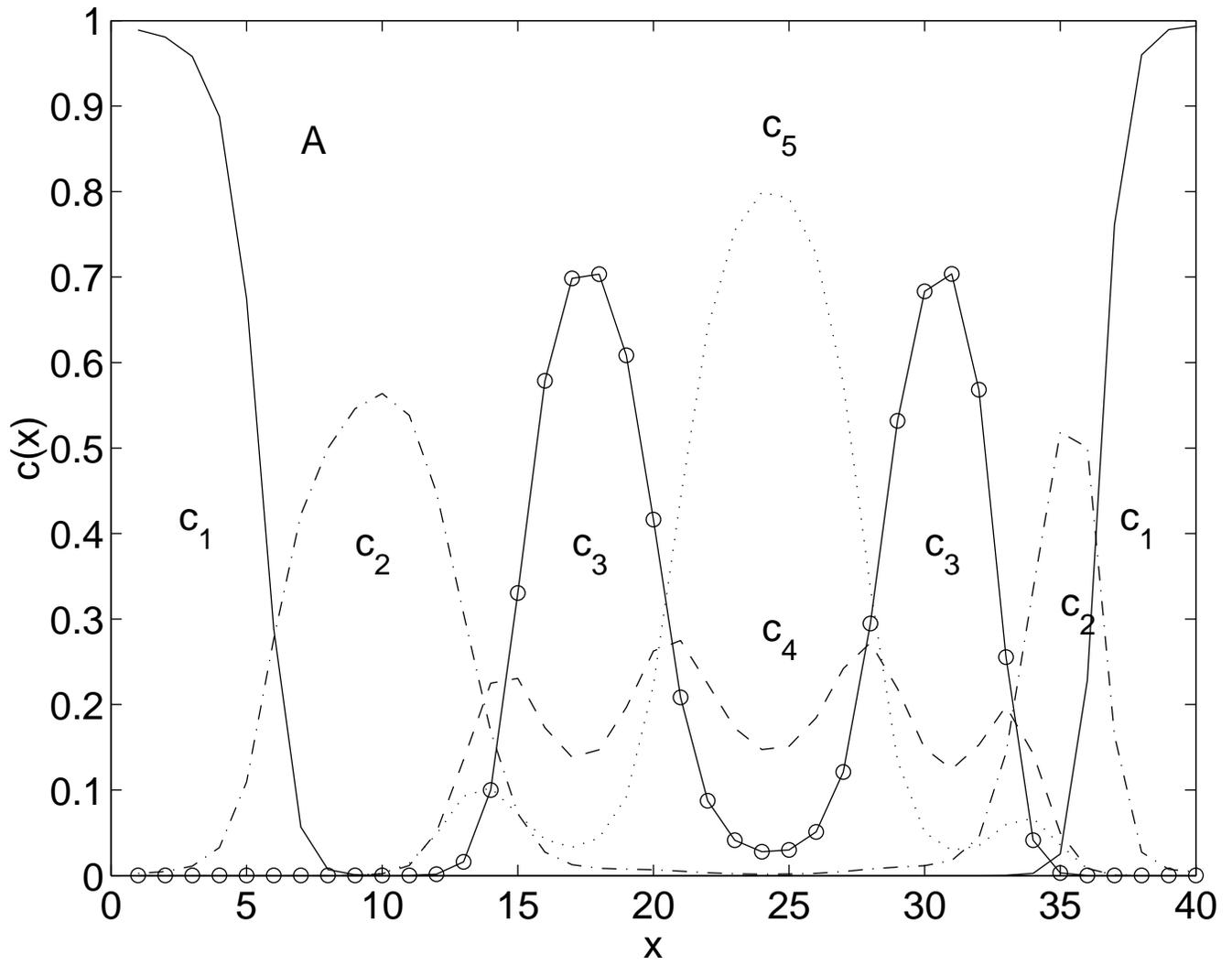
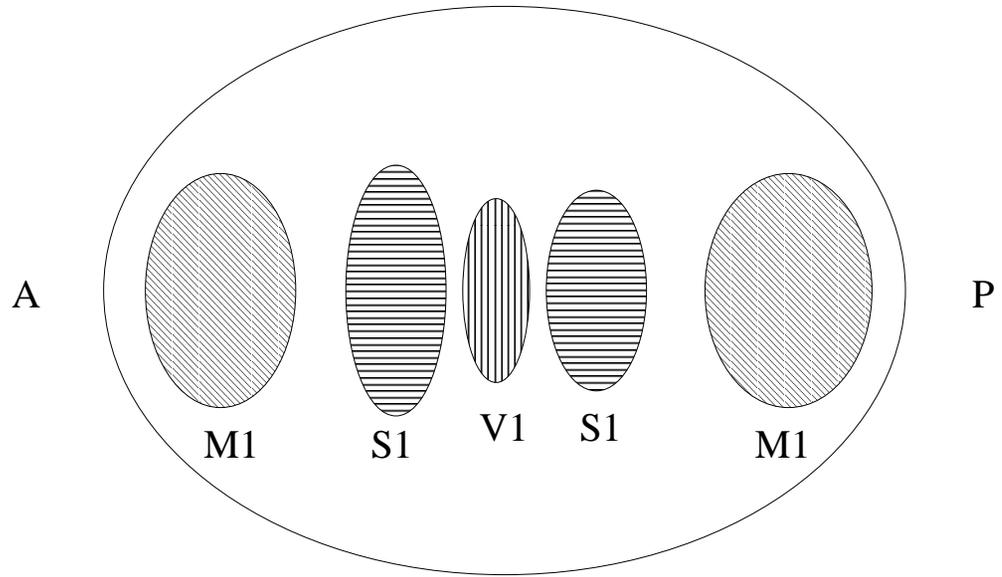
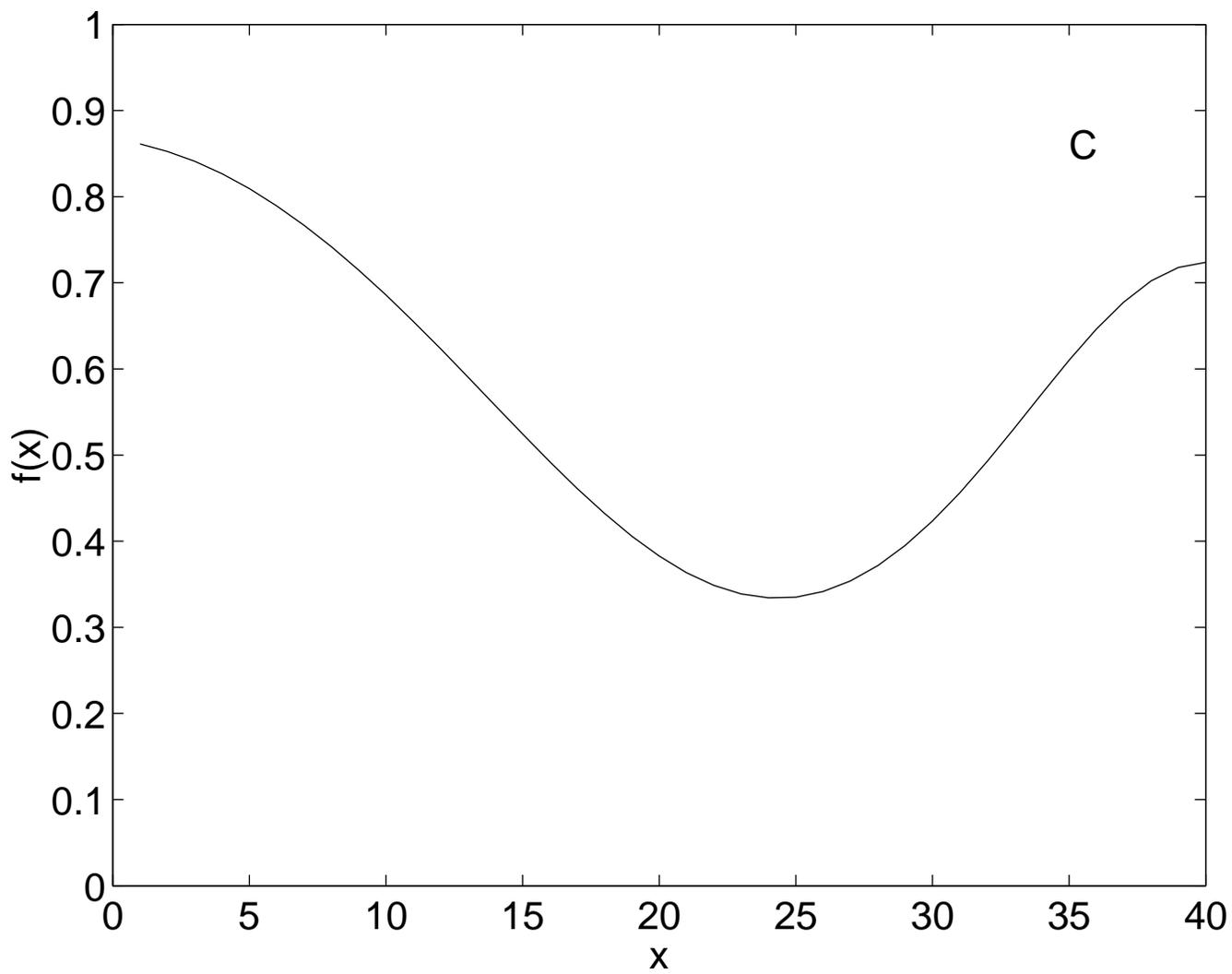
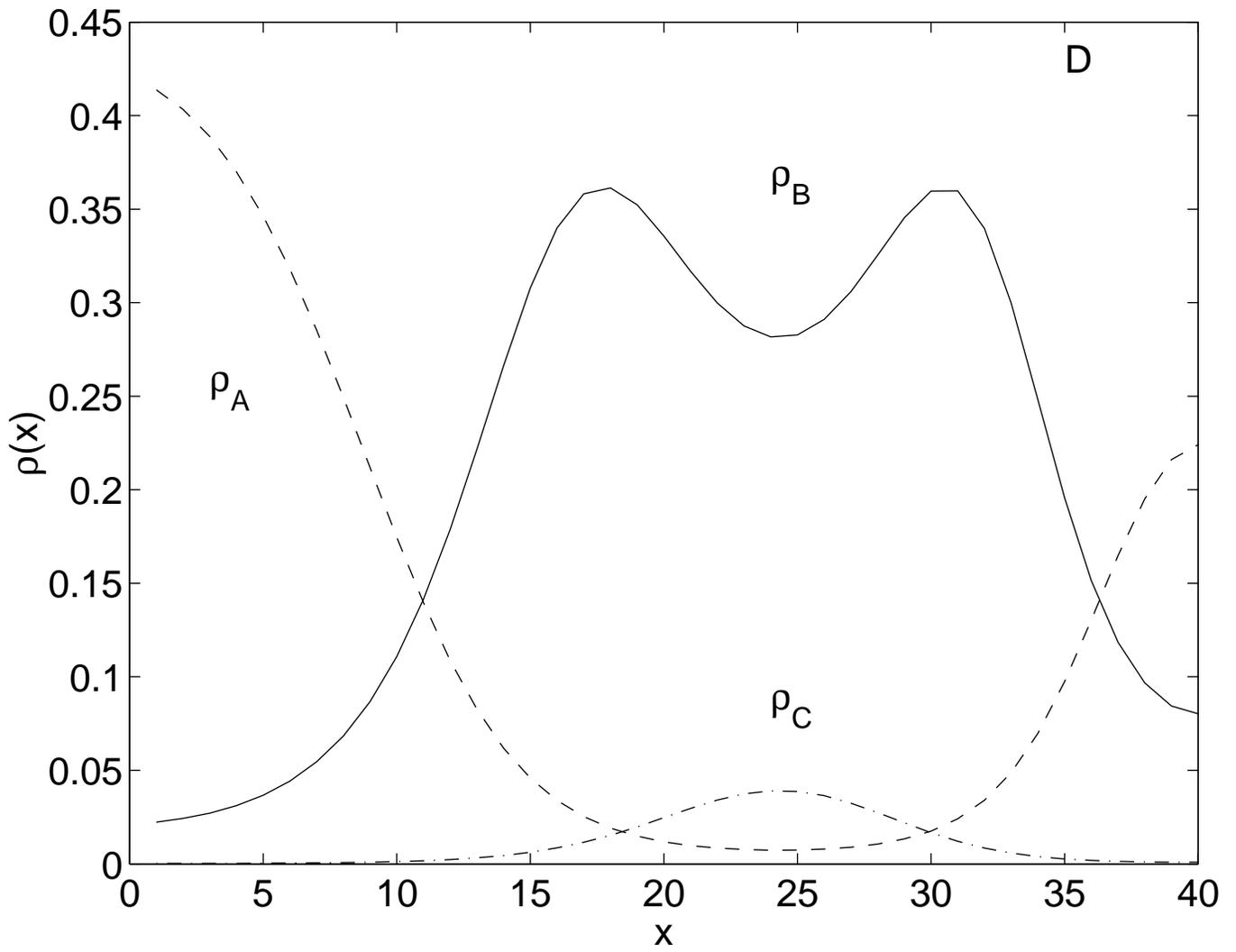
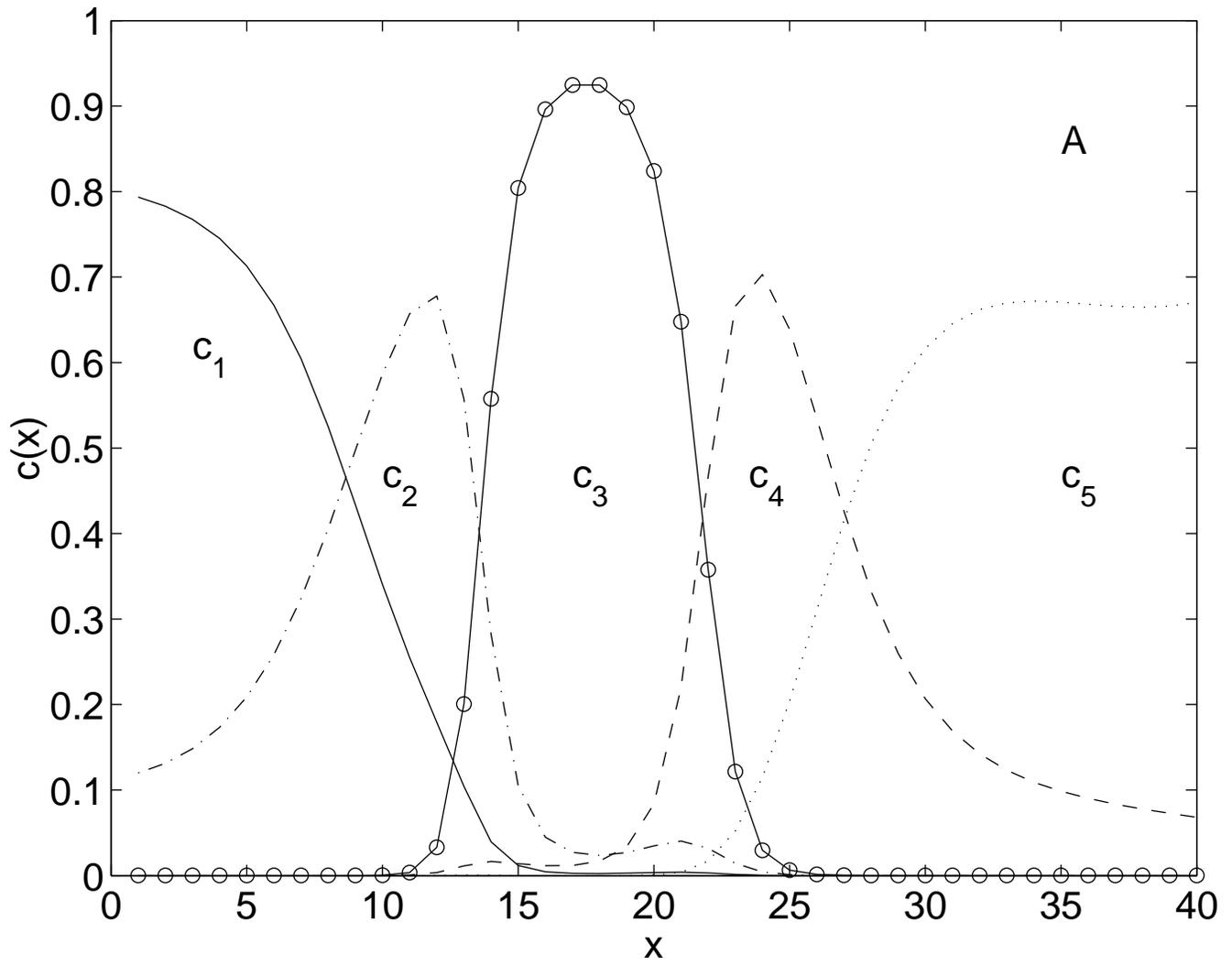
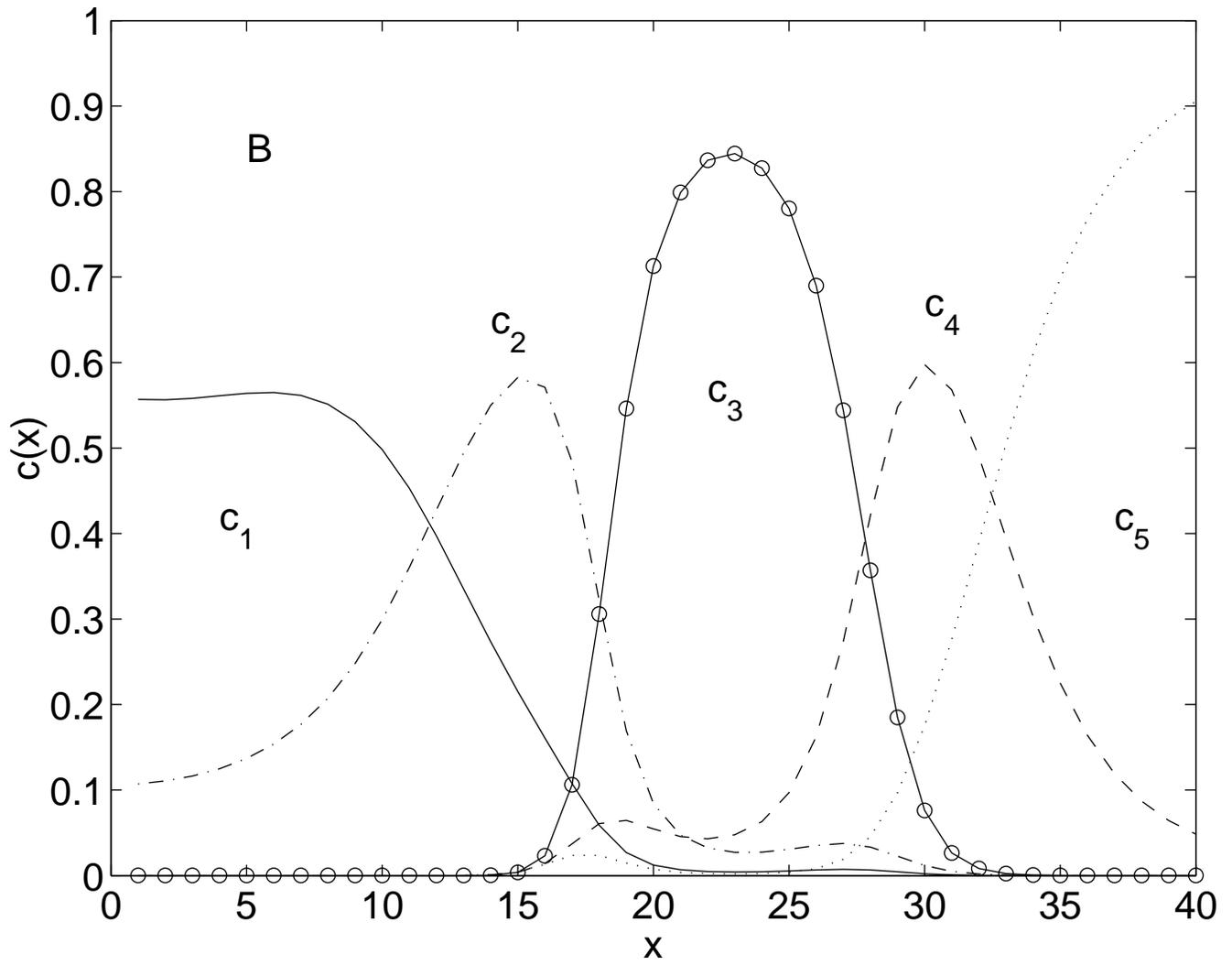
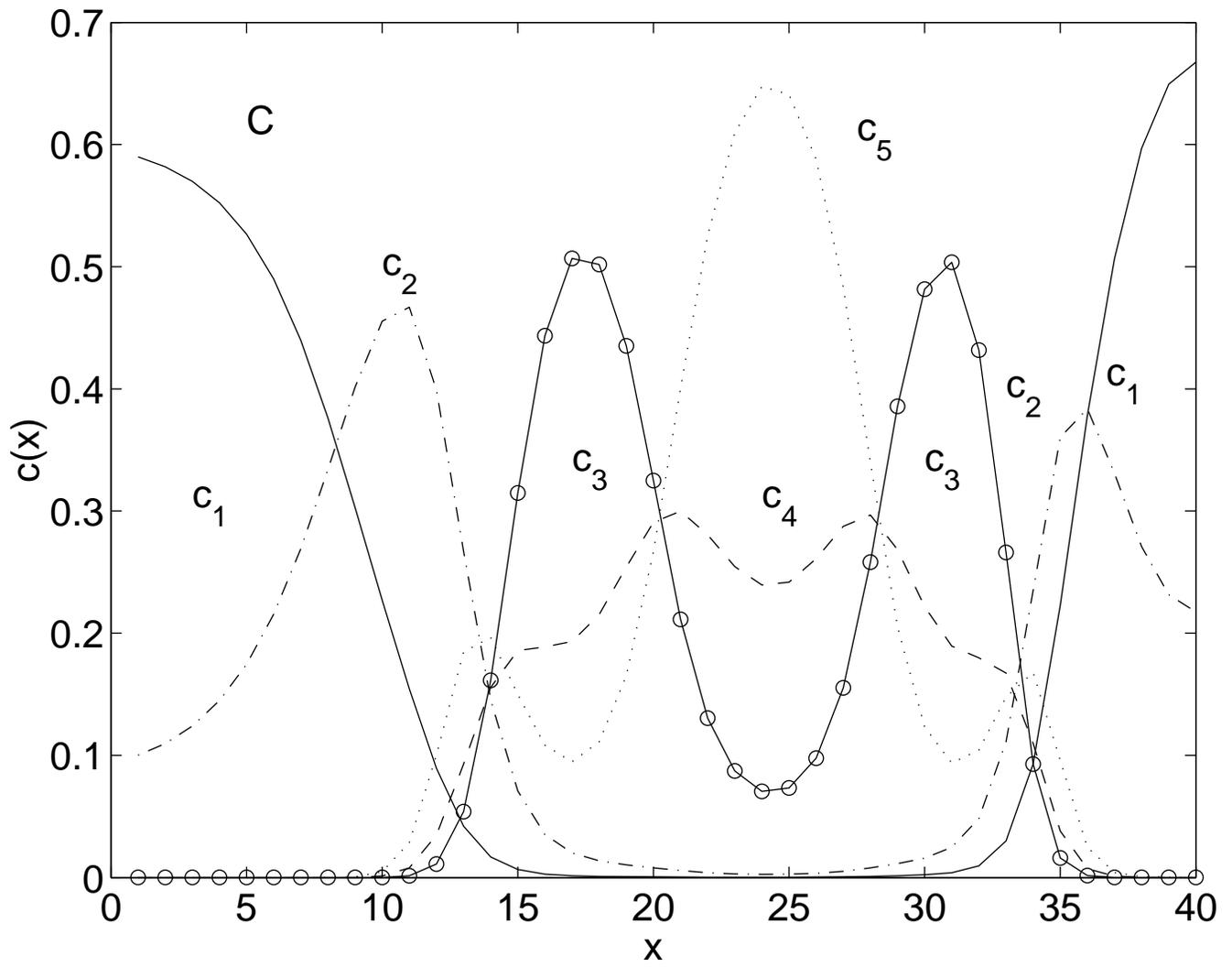
\end{document}